\documentclass[aps,twocolumn,nofootinbib,preprintnumbers,superscriptaddress]{revtex4-2}

\usepackage{color}
\usepackage{bbm}
\usepackage{tikz}
\usepackage{braket}
\usepackage{comment}

\usepackage[
pagebackref=false,
colorlinks=true,
linkcolor=blue,
urlcolor=blue,
filecolor=black,
citecolor=red,
pdfstartview=FitV,
pdftitle={},
pdfauthor={},
pdfsubject={},
pdfkeywords={},
pdfpagemode=None,
bookmarksopen=true
]{hyperref}

\usepackage[normalem]{ulem}
\usepackage{amsmath, amssymb, bm}
\usepackage{enumerate}
\usepackage{amsfonts}
\usepackage{epsfig}
\usepackage{mathbbol}
\usepackage{mathrsfs}
\definecolor{darkgreen}{rgb}{0.0, 0.5, 0.0}
\usepackage{makecell}
\usepackage{subcaption}

\newcommand{\dg}{\dagger}
\newcommand{\ra}{\rightarrow}

\newcommand{\tr}{\textrm{Tr}}
\newcommand{\bo}{\mathbbm{1}}

\newcommand{\mS}{\mathcal{S}}
\newcommand{\mT}{\mathcal{T}}
\newcommand{\mN}{\mathcal{N}}
\newcommand{\mP}{\mathcal{P}}

\begin{document}

\title{\textbf{Operator fusion from wavefunction overlaps: \\ Universal finite-size corrections and application to Haagerup model
}}

\author{Yuhan Liu}
\affiliation{Kadanoff Center for Theoretical Physics, University of Chicago, Chicago, IL~60637, USA}

\altaffiliation{These authors contributed equally to this work}

\author{Yijian Zou}
\affiliation{Stanford Institute for Theoretical Physics, Stanford University, Palo Alto, CA 94305, USA}
\altaffiliation{These authors contributed equally to this work}

\author{Shinsei Ryu}
\affiliation{ Department of Physics, Princeton University, Princeton, New Jersey, 08540, USA}
\begin{abstract}
    Given a critical quantum spin chain described by a conformal field theory (CFT) at long distances, it is crucial to understand the universal conformal data. One most important ingredient is the operator product expansion (OPE) coefficients, which describe how operators fuse into each other. It has been proposed in [Zou, Vidal, Phys. Rev. B \textbf{105}, 125125] that the OPE coefficients can be computed from overlaps of low-energy wavefunctions of the spin chain. 
    In this work, we establish that all conformal data including central charge, conformal dimensions, and OPE coefficients are encoded in the wavefunction overlaps, with universal finite-size corrections that depend on the operator content of the cyclic orbifold CFT. Thus this method allows us to numerically compute all the conformal data based solely on the low-energy eigenstates. The predictions are verified in the Ising and XXZ model. 
    As an application, we study the recently proposed Haagerup model built from the Haagerup fusion category. We find that the CFT has central charge $c\approx 2.1$ and the lowest spin-1 operator in the twisted sector has scaling dimension $1<\Delta_{J}\leq 1.4$. 
\end{abstract}

\maketitle
\tableofcontents

\section{Introduction}

Universality is one of the most prominent features of phase transitions \cite{wilson_renormalization_1974}. Near a continuous phase transition, physics at low energies is determined by the universal data. In 1+1 dimensions, a continuous phase transition is usually described by a conformal field theory (CFT) \cite{di_francesco_conformal_2012}. A CFT is completely determined by the conformal data, which consists of a set of primary operators $\phi_{\alpha}$ with scaling dimensions $\Delta_{\alpha}$ and conformal spins $s_{\alpha}$, the operator product expansion coefficients $C_{\alpha\beta\gamma}$ and the central charge $c$ \cite{belavin_infinite_1984,friedan_conformal_1984}.


Given a quantum critical spin chain, it is both important and challenging to compute the universal data. One approach, initiated by Cardy, Affleck and others in the 80's \cite{cardy_conformal_1984,blote_1986,affleck_universal_1986}, is based on the operator-state correspondence. In this approach, one starts with the low-energy eigenstate $|\phi_{\alpha}\rangle$ of the critical quantum spin chain with periodic boundary conditions (PBC). Once the Hamiltonian is properly normalized, the scaling dimensions $\Delta_{\alpha}$ and conformal spins $s_{\alpha}$ are given by the energies $E_{\alpha}$ and momenta $p_{\alpha}$ of the states. In order to compute the OPE coefficients, one has to first identify lattice operators that correspond to CFT primary operators \cite{cardy_operator_1986,Zou_2019}, which is in general hard to do without prior knowledge of the CFT. 

To overcome this difficulty more recently, it has been proposed in Refs.~\cite{zou2021universal,2021arXiv210809366Z} that the OPE coefficients can be numerically extracted with the eigenstates alone. The key is to start with three normalized
eigenstates $|\phi_{\alpha}\rangle$, $|\phi_{\beta}\rangle$, $|\phi_{\gamma}\rangle$ 
of the critical quantum spin chain with PBC and sizes $N_1,N_2$ and $N_3=N_1+N_2$, respectively. 
It is then shown that the wavefunction overlap 
\begin{align}
A_{\alpha\beta\gamma} = \langle \phi_{\gamma}| \phi_{\alpha}\phi_{\beta}\rangle
\end{align}
is related to the OPE coefficient $C_{\alpha\beta\gamma}$. This provides a numerically feasible way to extract the OPE coefficients by computing the wavefunction overlaps.

However, two important problems remain unsolved in Ref.~\cite{2021arXiv210809366Z}. 
Firstly, the paper only considered the case where $N_1=N_2$. It remains a question if OPE coefficients can still be extracted when the sizes are not equal. Secondly and more importantly, the relation between wavefunction overlap $A_{\alpha\beta\gamma}$ and the OPE coefficient $C_{\alpha\beta\gamma}$ 
is only expected to hold in the thermodynamic limit. 
At finite sizes, the finite-size corrections to $A_{\alpha\beta\gamma}$ may be large. 
Unlike the finite-size corrections to the energies $E_{\alpha}$, whose exponents are determined by the scaling dimensions of irrelevant operators of the CFT through standard conformal perturbation theory \cite{Henkel1999}, 
an understanding of finite-size corrections to $A_{\alpha\beta\gamma}$ is still missing. 
This makes it hard to extrapolate the finite-size data to the thermodynamic limit. 

In this work, we solve these two problems. 
First, using a generalized conformal map from three-sided cylinder to the complex plane which appeared in the string field theory literature \cite{1973NuPhB..64..205M, 2009JHEP...12..010B, SFT}, 
we establish a relation between wavefunction overlaps $A_{\alpha\beta\gamma}$ and OPE coefficients $C_{\alpha\beta\gamma}$ for general sizes. 
Second, we show that the exponents in the finite-size correction to $A_{\alpha\beta\gamma}$ are universal by mapping the wavefunction overlaps to three-point correlation functions of the cyclic orbifold of the original CFT. The exponents are determined by the operator content and fusion rules of the cyclic orbifold, which are in turn determined by the original CFT. The result also implies similar universal finite-size corrections in Renyi entropies of CFT, as observed in Ref.~\cite{Ohmori_2015}. As a benchmark, we examine the finite-size corrections to $A_{\alpha\beta\gamma}$ for the Ising and XXZ spin chain. All numerical data agrees with the theoretical predictions in high precision. 

Our result enables us to compute the conformal data from $A_{\alpha\beta\gamma}$ in a systematic way, with a proper estimation of finite-size corrections. 
For example, 
when $N_1 = N_2 = N$, 
the overlap $\langle \phi_{\gamma}|\bo\bo \rangle$ scales as
\begin{equation}
    A_{\bo\bo\gamma} = O( N^{-\frac{c}{8}-\frac{\Delta_{\gamma}}{2}}),
\end{equation}
where $|\bo\rangle$ is the ground state. The exponent gives the central charge and scaling dimensions without proper normalization of the lattice Hamiltonian. Another example is that if the CFT has a spin-1 conserved current $J$ and its conjugate $J^{*}$, then
\begin{equation}
\label{eq:ov_current}
    \frac{A_{JJ^{*}\bo}}{A_{\bo\bo\bo}} =\frac{1}{16}+O(N^{-2}).
\end{equation}
As an application, we compute the overlaps in a recently proposed Haagerup model \cite{huang2021numerical,vanhove2021critical}, which is conjectured to have $c=2$ and a spin-1 current $J$. Our numerical result, however, indicates that the CFT has $c\approx 2.1$ and that the proposed operator $J$ cannot be a conserved current that saturates the unitarity bound $\Delta_{J}\geq 1$. Rather, our numerical result supports that the operator is a spin-1 primary operator close to the unitarity bound, with an upper bound of the scaling dimension $\Delta_J\leq 1.4$.
 
Our work is both of interest from numerical and theoretical perspectives. Numerically, we provide a simple method to compute the conformal data based solely on the low-energy wavefunctions. Compared with the prior work, our method does not rely on input such as identification of lattice operators with CFT operators, symmetry of the model or correct normalization of the Hamiltonian, which makes it practical to explore a less-known CFT.  
Theoretically, we show that the wavefunction overlaps encode the
information of the corresponding cyclic orbifold CFT. It is therefore interesting
to investigate how to relate other orbifold CFT data with measurable quantities in the original CFT model.
Furthermore, as noted in Ref.~\cite{2021arXiv210809366Z}, the overlaps considered here can be regarded as wavefunctions of a toy three-sided wormhole in holography \cite{Balasubramanian_2014}. Thus it may be desirable to find a holographic interpretation of our results.


The rest of the paper is structured as follows. 
In Sec.~\ref{sec:overlap_main} we review the operator-state correspondence and derive the relation between wavefunction overlaps $A_{\alpha\beta\gamma}$ and the OPE coefficients $C_{\alpha\beta\gamma}$. 
In Sec.~\ref{sec:fss_main} we review the cyclic orbifold CFT 
and relate its operator content to the exponents in the universal finite-size corrections to $A_{\alpha\beta\gamma}$. 
As a corollary, 
we also derive finite-size corrections to the second Renyi entropy of CFT. 
In Sec.~\ref{sec:Ising_main} we compute $A_{\alpha\beta\gamma}$ for the Ising and XXZ model and verify that the finite-size corrections all match the analytical predictions with good accuracy. 
In Sec.~\ref{sec:haagerup_main} we apply our method to the Haagerup model and extract part of the conformal data. Finally, we discuss possible applications and extensions of the present work. 

\section{Operator fusion from Wavefunction overlaps}
\label{sec:overlap_main}
In this section we derive the relation between wavefunction overlaps and OPE coefficients of the CFT. We first represent the wavefunction overlaps as path integral on a three-sided cylinder. Then we use a conformal transformation to map the path integral to a three-point correlation function on the complex plane. Our derivation generalizes that of Ref.~\cite{2021arXiv210809366Z} in that we do not restrict two sides of the cylinder to have the same size. 
\subsection{CFT on a cylinder}
Consider a CFT on a Euclidean cylinder with circumference $L$. The compact dimension represents the space with coordinate $x\in [0,L)$, and the non-compact dimension represents the imaginary time with coordinate $\tau \in (-\infty,\infty)$. It is convenient to use the complex coordinates, denoted as $z = \tau+ix$ and $\bar{z} = \tau-ix$. The Hilbert space is supported on the equal-time slices.
A CFT is composed of a set of local scaling operators that are covariant under global rescaling of the spacetime, where the eigenvalue is the scaling dimension. The scaling operators are organized into conformal towers. For each tower, there is a primary operator $\phi_{\alpha}$ with the lowest scaling dimension $\Delta_{\alpha}$, and other descendant operators (later abbreviated as des.) with scaling dimensions that differ $\Delta_{\alpha}$ by integers. 

A crucial property of the CFT is the operator-state correspondence, which states that each scaling operator corresponds to a state on the cylinder. The energies of the states are related to the scaling dimensions by
\begin{equation}
    E_{\alpha} = \frac{2\pi}{L} \left(\Delta_{\alpha} - \frac{c}{12}\right),
\end{equation}
where $c$ is the central charge of the CFT. For a unitary CFT, the ground state $|\bo\rangle$ corresponds to the identity operator $\bo$ with scaling dimension $\Delta_{\bo} = 0$. Other states can be created by acting with the scaling operator on the ground state at infinity,
\begin{eqnarray}
\label{eq:op_state1}
    |\phi_{\alpha}\rangle &=& \phi_{\alpha}(-\infty) |\bo\rangle. \\
\label{eq:op_state2}
    \langle\phi_{\alpha}| &=& \langle \bo| \phi^{*}_{\alpha}(+\infty),
\end{eqnarray}
where
\begin{eqnarray}
\label{eq:phiinf}
    \phi_{\alpha}(\pm \infty) \equiv \left(\frac{2\pi}{L}\right)^{-\Delta_{\alpha}} \lim_{\tau\rightarrow \pm\infty} e^{\pm \frac{2\pi}{L}\Delta_{\alpha}\tau} \phi_{\alpha}(\tau,0),
\end{eqnarray}
is a shorthand notation and $\phi^{*}_{\alpha}$ is the conjugation of $\phi_{\alpha}$. In many cases the primary operator is self-conjugate and we will not distinguish the two. The states form an orthonormal basis, 
\begin{equation}
    \langle \phi_{\alpha}|\phi_{\beta}\rangle = \delta_{\alpha\beta}.
\end{equation}
Given three primary operators $\phi_{\alpha},\phi_{\beta},\phi_{\gamma}$, the three-point correlation function is determined by the operator product expansion (OPE), where the only independent coefficient is given by
\begin{equation}
\label{eq:OPEcyl}
    C_{\alpha\beta\gamma} = \left( \frac{2\pi}{L}
    \right)^{-\Delta_{\alpha}} \langle \phi_{\gamma}|\phi_{\alpha}(0)|\phi_{\beta}\rangle.
\end{equation}
Other correlation functions, including those of descendant operators, are completely determined by the OPE coefficients of primary operators. If $C_{\alpha\beta\gamma}\neq 0$, there is a fusion channel $\phi_{\alpha} \times \phi_{\beta} \rightarrow \phi_{\gamma}$, and we denote it by $\mN_{\alpha,\beta,\gamma} \neq 0$. Otherwise, the fusion channel is said to be forbidden, and $\mN_{\alpha,\beta,\gamma} = 0$. In all cases considered in this work, $\mN_{\alpha,\beta,\gamma}$ is either $0$ or $1$, depending on whether the fusion channel is forbidden or allowed. 

\subsection{Path integral for wavefunction overlaps}
\label{sec:wavefunction_overlap}

\begin{figure}
    \centering
    \includegraphics[width=0.7\linewidth]{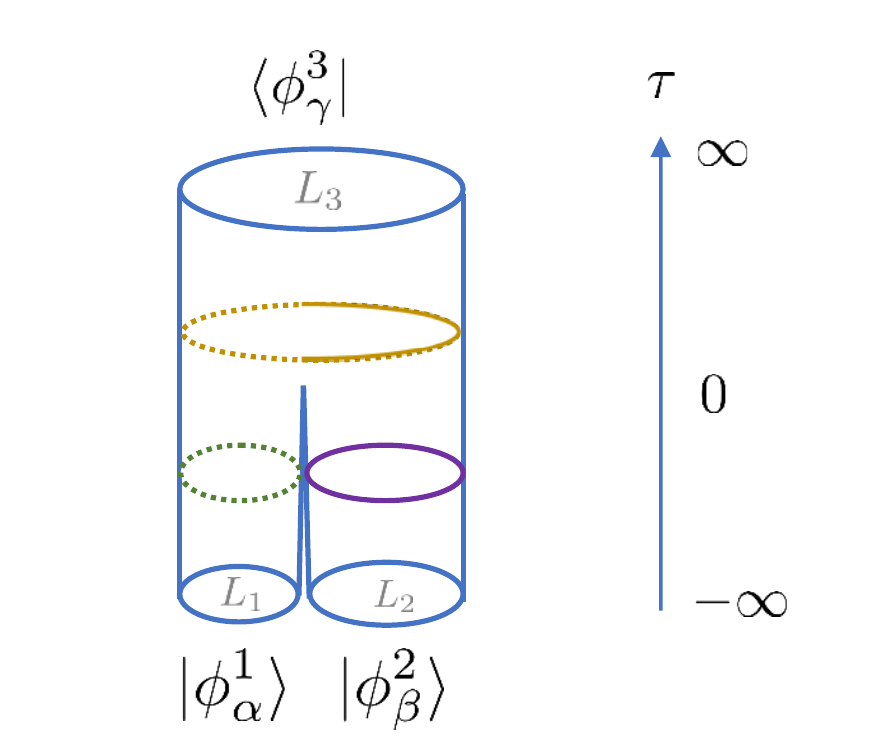}
    \caption{The path integral for the wavefunction overlap $A_{\alpha\beta\gamma}=\langle \phi^{3*}_{\gamma}|\phi^{1}_{\alpha}\phi^{2}_{\beta}\rangle$. The geometry is a three-sided cylinder $\Sigma$ with circumference $L_1+L_2=L_3$. }
    \label{fig:3side}
\end{figure}

\begin{figure}
    \begin{subfigure}[b]{\linewidth}
     \centering
     \includegraphics[width=\textwidth]{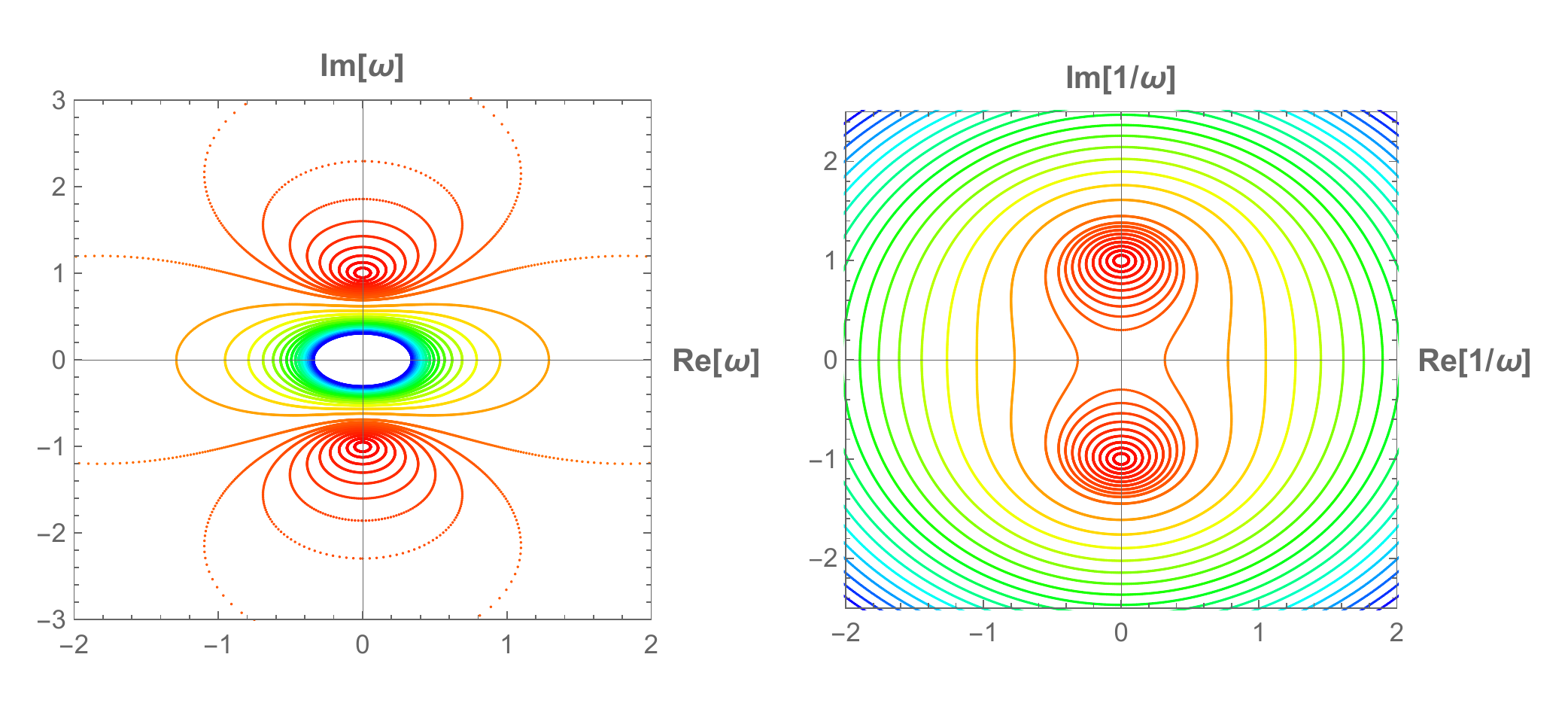}
     \caption{}
\end{subfigure}
\begin{subfigure}[b]{\linewidth}
     \centering
     \includegraphics[width=\textwidth]{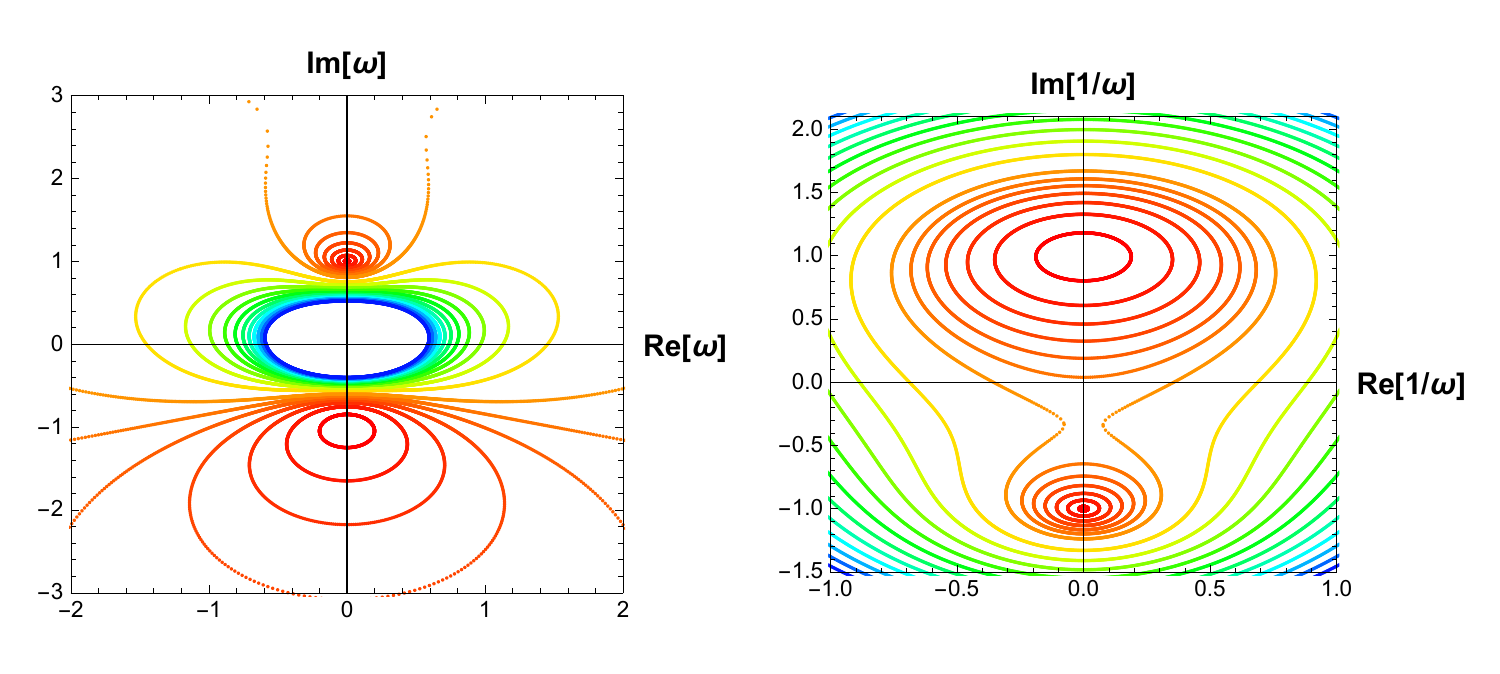}
     \caption{}
\end{subfigure}
    \caption{Illustration of the conformal transformations that map three-sided cylinder to the complex plane. (a) Eq.~\eqref{eq:conf_trans1} for the case of $L_1=L_2$ and (b) Eq.~\eqref{eqn:third}, for the case of $L_2=2L_1$. Colored lines represent equal time slices with equal real part of $z$. Darker color corresponds to larger $\tau$. Red lines correspond to equal time slices on cylinders $1$ and $2$, and the blue lines correspond to equal time slices on cylinder $3$. For better illustration purposes we show both $w$ and $1/w$ for the two conformal transformations.
    }
    \label{fig:second_trans}
\end{figure}

Let us consider three cylinders with circumference $L_1$,$L_2$ and $L_3$ with $L_1+L_2=L_3$. On each cylinder a primary operator is inserted at infinity to create a state. Then the wavefunction overlaps can be represented by a path integral on the three-sided cylinder $\Sigma$ with three insertions 
\begin{equation}
\begin{aligned}
    {A_{\alpha\beta\gamma}=}&\langle \phi^{3}_{\gamma}|\phi^{1}_{\alpha}\phi^{2}_{\beta}\rangle\\
    =& \int_{\Sigma} D\Phi\, \phi^{1}_{\alpha}(-\infty) \phi^{2}_{\beta}(-\infty) \phi^{3*}_{\gamma}(+\infty) e^{-S_{\mathrm{CFT}}[\Phi]},
\end{aligned}
\end{equation}
where $D\Phi$ is the functional measure and $S_{\mathrm{CFT}}$ is the action of the CFT,
\begin{equation}
    S_{\mathrm{CFT}}[\Phi] = \int_{\Sigma} d\tau dx \, \mathcal{L}_{\mathrm{CFT}}[\Phi], 
\end{equation}
where $\mathcal{L}_{\mathrm{CFT}}[\Phi]$ is the Lagrangian density. An illustration of the three-sided cylinder is shown in Fig.~\ref{fig:3side}.
The partition function is the path integral with no insertions, thus
\begin{equation}
     \int_{\Sigma} D\Phi\,  e^{-S_{\mathrm{CFT}}[\Phi]} = \langle \bo^{3}|\bo^{1}\bo^{2}\rangle.
\end{equation}
Taking the ratio gives
\begin{equation}
    \frac{\langle \phi^{3}_{\gamma}|\phi^{1}_{\alpha}\phi^{2}_{\beta}\rangle}{\langle \bo^{3}|\bo^{1}\bo^{2}\rangle} = \langle \phi^{1}_{\alpha}(-\infty) \phi^{2}_{\beta}(-\infty) \phi^{3*}_{\gamma}(+\infty) \rangle_{\Sigma}
\end{equation}
Note that the three cylinders meet at $z_0 = iL_1$ on the $\tau=0$ time slice.

Next, we use a conformal transformation $w(z)$ to map $\Sigma$ to the complex plane ${\mathbbm{C}}$. Inspired by the string field theory literature \cite{1973NuPhB..64..205M, 2009JHEP...12..010B}, the conformal map is defined implicitly by 
\begin{align}
&
   z+\lambda = 
   \nonumber \\
   &
   \frac{1}{2\pi}[L_1 \log(w-w_1) + L_2 \log(w-w_2) -L_3 \log(w-w_3)],
\end{align}
where $w_1,w_2$ and $w_3$ are arbitrary complex numbers and $\lambda$ depends on the $w$'s. The role of $\lambda$ will be evident shortly. It is clear that the conformal transformation maps the three infinity points to $w_1$, $w_2$ and $w_3$, respectively. One example considered in Ref.~\cite{2021arXiv210809366Z} is when $L_3=2L_1=2L_2=L$, $w_1=i, w_2=-i,w_3=0$ and $\lambda=0$, where the conformal transformation simplifies to an explicit form
\begin{equation}
\label{eq:conf_trans1}
    w = \frac{1}{\sqrt{e^{4\pi z/L}-1}}.
\end{equation}
There is a second-order algebraic branch point at $z_0=0$ where the three cylinders meet. 
This is illustrated in Fig.~\ref{fig:second_trans} (a). For general choices of parameters, the conformal transformation $w(z)$ is hard to express explicitly. However, the conformal transformation still possesses a second-order algebraic branch point, which can be shifted to $z_0=0$ by tuning $\lambda$. Let $w(z_0) = w_0$, then the branch-point condition demands that
\begin{equation}
    \frac{dz}{dw}(w_0) = 0.
\end{equation}
Solving the equation gives $w_0$. Substituting $w_0$ back to $w(z_0) = w_0$ then fixes $\lambda$. For example, if we choose $L_2=2L_1=2L/3$ and $w_1=i,w_2=-i,w_3=0$, then $w_0=3i$ and $\lambda=(L\log (32/27))/(6\pi)$. The conformal transformation then becomes
\begin{eqnarray}
    \frac{32}{27}\cdot e^{\frac{6\pi z}{L}}=\frac{(w-i)(w+i)^2}{w^3},
     \label{eqn:third}
\end{eqnarray}
which is illustrated in Fig.~\ref{fig:second_trans} (b).

Using conformal transformation of primary operators, the correlation function on $\Sigma$ transforms into a three-point correlation function on the complex plane $\mathbbm{C}$,
\begin{eqnarray}
    &~&\langle \phi^{1}_{\alpha}(-\infty) \phi^{2}_{\beta}(-\infty) \phi^{3*}_{\gamma}(+\infty) \rangle_{\Sigma} \nonumber \\ 
    \label{eq:corr3_Sigma}
    &=&  \prod_i\left|J_i\right|^{-\Delta_{i}} \langle \phi^{1}_{\alpha}(w_1) \phi^{2}_{\beta}(w_2) \phi^{3*}_{\gamma}(w_3) \rangle_{{\mathbbm{C}}},
\end{eqnarray}
where 
\begin{equation}
    J_i = e^{\pm\frac{2\pi}{L_i}\lambda}\frac{dz}{dw}(w_i)
\end{equation}
is the conformal factor. The prefactor $e^{\pm\frac{2\pi}{L_i}\lambda}$ comes from the definition Eq.~\eqref{eq:phiinf}, where the plus sign is taken for $\phi^{3}_{\gamma}$ and the minus sign is taken for $\phi^{1}_{\alpha}$ and $\phi^{2}_{\beta}$.
The three-point correlation function on the complex plane has the standard form
\begin{equation}
\label{eq:corr3_Z}
    \langle \phi^{1}_{\alpha}(w_1) \phi^{2}_{\beta}(w_2) \phi^{3*}_{\gamma}(w_3) \rangle_{{\mathbbm{C}}} = C_{\alpha\beta\gamma} \prod_{i<j} |w_i-w_j|^{-d_{ij}},
\end{equation}
where $d_{12} = \Delta_{\alpha}+\Delta_{\beta}-\Delta_{\gamma}$, and others follow from permutation symmetry.

Finally, combining Eqs.~\eqref{eq:corr3_Sigma}, \eqref{eq:corr3_Z} and the expression for $\lambda$, we obtain
\begin{eqnarray}
    &~&\frac{\langle \phi^{3}_{\gamma}|\phi^{1}_{\alpha}\phi^{2}_{\beta}\rangle}{\langle \bo^{3}|\bo^{1}\bo^{2}\rangle} \nonumber \\ 
    \label{eq:ov3state_gen}
    &=& \left[\left(\frac{L_3}{L_1}\right)^{\frac{L_1}{L_3}} \left(\frac{L_3}{L_2}\right)^{\frac{L_2}{L_3}}\right]^{-\frac{L_3}{L_1}\Delta_{\alpha}-\frac{L_3}{L_2}\Delta_{\beta}+\Delta_{\gamma}} C_{\alpha\beta\gamma}.
    \label{eqn:A_full_expression}
\end{eqnarray}
The result is independent of the parameters $w_1,w_2$ and $w_3$, as expected. The special case in Ref.~\cite{2021arXiv210809366Z} corresponds to $L_3=2L_1=2L_2$, where the expression simplifies to
\begin{equation}
\label{eq:ov3state}
    \frac{\langle \phi^{3}_{\gamma}|\phi^{1}_{\alpha}\phi^{2}_{\beta}\rangle}{\langle \bo^{3}|\bo^{1}\bo^{2}\rangle} = 2^{-2\Delta_{\alpha}-2\Delta_{\beta}+\Delta_{\gamma}} C_{\alpha\beta\gamma}.
\end{equation}

On a finite lattice, Eqs.~\eqref{eq:ov3state_gen} and \eqref{eq:ov3state} are satisfied up to finite-size corrections. 
Physically, the finite-size corrections come from the singular point where the three cylinders meet. 
In order to define the geometry, one has to put a UV cutoff $\epsilon$ around the singular point, analogous to the UV cutoff in the calculation of entanglement entropy in CFT \cite{Holzhey1994,Calabrese_2004,Calabrese_2009}. 
On the lattice, the UV cutoff $\epsilon$ is 
inversely proportional to the number of spins $N$. So far the conformal transformation works in the limit of $\epsilon\rightarrow 0$. 
Thus we expect that Eqs.~\eqref{eq:ov3state_gen} and \eqref{eq:ov3state} strictly hold 
in the thermodynamic limit $N\rightarrow\infty$. 
In what follows, we focus on the special case with $L_1 = L_2$ and study the finite-size corrections to Eq.~\eqref{eq:ov3state}.
To this end, 
it is useful to switch 
from a path integral of a CFT on 
$\Sigma$ to an equivalent viewpoint,
the $\mathbb{Z}_2$ orbifold of the CFT
defined on a cylinder.




\section{Universal finite-size corrections}
\label{sec:fss_main}
In this section we consider finite-size correction to Eq.~\eqref{eq:ov3state}.
Specifically,
we consider the normalized 
wavefunction overlap
\begin{align}
\tilde{A}_{\alpha\beta \gamma}    
\equiv
    \frac{\langle \phi^{3}_{\gamma}|\phi^{1}_{\alpha}\phi^{2}_{\beta}\rangle}
    {\langle \bo^{3}|\bo^{1}\bo^{2}\rangle} 
\end{align}
and its finite-size corrections
\begin{equation}
\label{eq:FSS}
\tilde{A}_{\alpha\beta \gamma}    
    = 
    \tilde{A}^{(0)}_{\alpha\beta\gamma} + \sum_{p_{\alpha\beta\gamma}>0} \tilde{A}^{(p)}_{\alpha\beta\gamma} N^{-p_{\alpha\beta\gamma}}, \end{equation}
where 
\begin{equation}
    \tilde{A}^{(0)}_{\alpha\beta\gamma} = 2^{-2\Delta_{\alpha}-2\Delta_{\beta}+\Delta_{\gamma}} C_{\alpha\beta\gamma}.
\end{equation}
We show that $p_{\alpha\beta\gamma}$ only take on specific values determined by the scaling dimensions and fusion rules of the cyclic orbifold of the original CFT.
They are given explicitly in Eq.~\eqref{eq:p_set}. 
Cyclic orbifold has played an important role in the study of entanglement in quantum field theories. It is defined by gauging the cyclic permutation group $\mathbb{Z}_\mathsf{N}$ of $\mathsf{N}$ copies of the original CFT. In this work we focus on $\mathsf{N}=2$, but generalizations to larger $\mathsf{N}$ is straightforward. 

In the following, we first review the definition and the operator content of the cyclic orbifold. 
We then map the wavefunction overlaps to the OPEs of the cyclic orbifold and derive the universal finite-size corrections. Finally, we remark on the implication of our construction on the finite-size corrections to entanglement in CFT.
\subsection{Cyclic orbifold on the cylinder}
Given any CFT with Lagrangian density $\mathcal{L}_{\mathrm{CFT}}[\Phi]$, we can define the cyclic orbifold as follows \cite{borisov1998systematic,klemm1990orbifolds}. We start with two copies of the original CFT known as the the mother CFT, whose Lagrangian density is given by
\begin{equation}
    \mathcal{L}_{\mathrm{mother}}[\Phi_1,\Phi_2] = \mathcal{L}_{\mathrm{CFT}}[\Phi_1]+\mathcal{L}_{\mathrm{CFT}}[\Phi_2].
\end{equation}
The mother CFT has a $\mathbb{Z}_2$ global symmetry that swaps the two copies. Denote the generator of the $\mathbb{Z}_2$ symmetry as $\sigma$, then
\begin{equation}
    \sigma \Phi_i= \Phi_{i+1},
\end{equation}
where the subscript is understood to have periodicity $2$. The cyclic orbifold theory is obtained by promoting the global symmetry to a gauge symmetry. The meaning of gauging a symmetry is twofold. Firstly, we enlarge the Hilbert space by allowing twisted boundary conditions,
\begin{equation}
    \Phi_i(\tau,x+L) = \sigma \Phi_i(\tau,x).
\end{equation}
This plays the role of gauge fields in usual gauge theory such as electrodynamics. Secondly, we project the Hilbert space onto subspace that is invariant under $\sigma$ 
\footnote{More precisely, we need to keep every inequivalent irreducible representation once in order to have a modular invariant partition function. 
This subtlety arises in cyclic orbifold.}. 
This amounts to enforcing gauge constraints in usual gauge theory. 

To see the operator content of the cyclic orbifold, first note that there are two sectors of operators, the untwisted sector and the twisted sector. The untwisted sector contains eigenstates of the mother CFT with periodic boundary conditions. The twisted sector contains eigenstates of the mother CFT with twisted boundary conditions. Second, only the operators that are invariant under the $\sigma$ swapping is kept. This is essential to obtain a consistent CFT because any operator that is not invariant under $\sigma$ is non-local with respect to the twist sector. In the following subsection, we review the operator content of cyclic orbifold in more detail.

\subsection{Operator content of cyclic orbifold}
\label{sec:cyclic_op}

As has been well studied, 
the operator content of the cyclic orbifold is completely determined by the original CFT.  The operator content is composed of the untwisted sector and the twisted sector. 
In the untwisted sector, 
the states are obtained from the tensor product theory. 
For $\alpha\neq \beta$, the symmetric states $|\phi_{\alpha,i}\rangle|\phi_{\beta,j}\rangle+|\phi_{\beta,j}\rangle|\phi_{\alpha,i}\rangle$ and anti-symmetric states $|\phi_{\alpha,i}\rangle|\phi_{\beta,j}\rangle-|\phi_{\beta,j}\rangle|\phi_{\alpha,i}\rangle$ in the Hilbert space are isomorphic. The labels $i,j$ in the above expressions are the labels for descendants. Due to the isomorphism, we only keep one of the copies, say, the symmetric states. We denote the orbifold primary operator corresponds to this copy as:
\begin{eqnarray}
     \phi_{(\alpha,\beta)},
\end{eqnarray}
with scaling dimension:
\begin{eqnarray}
     \Delta_{(\alpha,\beta)}=\Delta_\alpha+\Delta_\beta.
\end{eqnarray}

For $\alpha=\beta$, the above isomorphism fails because when $\alpha=\beta,i=j$, the antisymmetric state does not exist. For this reason, both the copies should be kept, and we denote the two orbifold primary operators as:
\begin{eqnarray}
     \phi_{(\alpha,\alpha)_s},\quad \phi_{(\alpha,\alpha)_a},
\end{eqnarray}
where $s$ stands for symmetric and $a$ stands for anti-symmetric. 
Their scaling dimensions are:
\begin{equation}
    \begin{aligned}
         &\Delta_{(\alpha,\alpha)_s}=2\Delta_\alpha,\\
         &\Delta_{(\alpha,\alpha)_a}=\begin{cases}
    2\Delta_\alpha+2, &\text{if $\phi_\alpha$ has level-1 des.}\\
    2\Delta_\alpha+4,              & \text{otherwise},
\end{cases}
    \end{aligned}
\end{equation}
where `des.' stands for descendant. 
Note that $\phi_{(\alpha,\alpha)_a}$ has larger scaling dimension than $\phi_{(\alpha,\alpha)_s}$, because 
antisymmetrization 
of 
the state with lowest $L_0$ eigenvalue $|\phi_\alpha\rangle|\phi_\alpha\rangle$ 
simply vanishes. 
Hence,  
a primary state 
$|\phi_{\alpha}\rangle$
must be combined with
its descendant and then 
antisymmetrized.
As far as 
the chiral 
algebra involves only
conformal symmetry,
only $\bo$ operator does not have a level-1 descendant. (Namely, $|\bo\rangle$ is the conformal vacuum.) This is no longer the case if the Kac-Moody algebra is taken into account, where the $\bo$ operator does have level-1 descendants which are current operators. The identity operator in the orbifold theory is $\phi_{(\bo,\bo)_s}$, which has $\Delta=0$.

Summarizing, if the original theory has $n$ primary operators, then the untwisted sector of the orbifold theory has $n(n-1)/2+2n = n(n+3)/2$ primary operators. In Appendix \ref{app:modular_trans} we list the characters of each orbifold primary.


In the twisted sector, the operator with the lowest scaling dimension is denoted as $\tau_{(\bo,\hat{0})}$, whose scaling dimension is completely determined by the central charge,
\begin{equation}
    \Delta_{(\bo,\hat{0})} =\frac{c}{8}.
\end{equation}
This is known as the branch-point twist operator in the literature. However, there are other primary operators in the twisted sector, which plays an important role in this work. One may fuse $\tau_{(\bo,\hat{0})}$ with each primary operator $\phi_{(\alpha,\bo)}$ in the untwisted sector to generate other twist operators:
\begin{eqnarray}
     \tau_{(\bo,\hat{0})}\times \phi_{(\alpha,\bo)}\ra \tau_{(\alpha,\hat{0})}+\tau_{(\alpha,\hat{1})},\quad \alpha\neq \bo.
\end{eqnarray}
For each $\alpha$, there are two primary operators in the twisted sector\footnote{This is also true for $\alpha=\bo$, and $\tau_{(\bo,\hat{1})}$ can be generated using channel $\tau_{(\beta,\hat{0})}\times \phi_{(\beta,\bo)}$, $\beta\neq\bo$.}:
\begin{eqnarray}
     \tau_{(\alpha,\hat{0})},\quad \tau_{(\alpha,\hat{1})},
\end{eqnarray}
with scaling dimensions:
\begin{equation}
    \begin{aligned}
         \Delta_{(\alpha,\hat{0})} &=\frac{c}{8}+\frac{\Delta_{\alpha}}{2}\\
         \Delta_{(\alpha,\hat{1})} &=\begin{cases}
         \frac{c}{8}+\frac{\Delta_{\alpha}}{2}+1,& \text{if $\phi_\alpha$ has level-1 des.}\\
         \frac{c}{8}+\frac{\Delta_{\alpha}}{2}+3,& \text{otherwise}. 
         \end{cases}
    \end{aligned}
\end{equation}
Summarizing, if the original theory has $n$ primary operators, then the orbifold theory has $2n$ primary operators in the twisted sector.

Finally, we need the fusion rules of the orbifold theory. It is intuitively clear that an operator in the untwisted sector fusing with an operator in the twisted sector only gives operators in the twisted sector. Furthermore, it has been shown \cite{borisov1998systematic} that the fusion rules are completely determined by the modular $\mS$ and $\mT$ matrices of the original CFT. For an introduction to modular matrices and details of the derivation, see Appendix  \ref{app:modular_trans}. We will need the following two fusion coefficients,
\begin{equation}
    \mN_{(\alpha,\beta),{(\gamma,\hat{\psi})},{(\delta,\hat{\chi})}}=\sum_{\eta} \frac{\mS_{\alpha\eta}\mS_{\beta\eta}\mS_{\gamma\eta}\mS^*_{\eta\delta}}{\mS_{\bo\eta}^2},
    \label{eqn:fusion_orbifold_1}
\end{equation}
and
\begin{equation}
\begin{aligned}
    &\mN_{(\alpha,\alpha)_s,{(\gamma,\hat{\psi})},{(\delta,\hat{\chi})}}=\\
    &\frac{1}{2}\sum_\eta \frac{\mS_{\alpha\eta}^2 \mS_{\gamma\eta}\mS^*_{\eta\delta}}{\mS^2_{\bo\eta}}+\frac{1}{2}e^{i\pi(\psi+\chi)}\sum_\eta\frac{\mS_{\alpha\eta}\mP_{\gamma\eta}\mP^*_{\eta,\delta}}{\mS_{\bo,\eta}},
\end{aligned}
\label{eqn:fusion_orbifold_2}
\end{equation}
where $\mP=\mT^{1/2}\mS \mT^2 \mS \mT^{1/2}$. In particular, Eq.~\eqref{eqn:fusion_orbifold_1} can be simplified to 
\begin{equation}
    \mN_{(\alpha,\beta),{(\gamma,\hat{\psi})},{(\delta,\hat{\chi})}} = \sum_{s} \mN_{\alpha,\beta,s} \mN_{\gamma,\delta,s^{*}} 
\end{equation}
if all primary fields are self-conjugate.

\begin{figure}
    \centering
    \includegraphics[width=\linewidth]{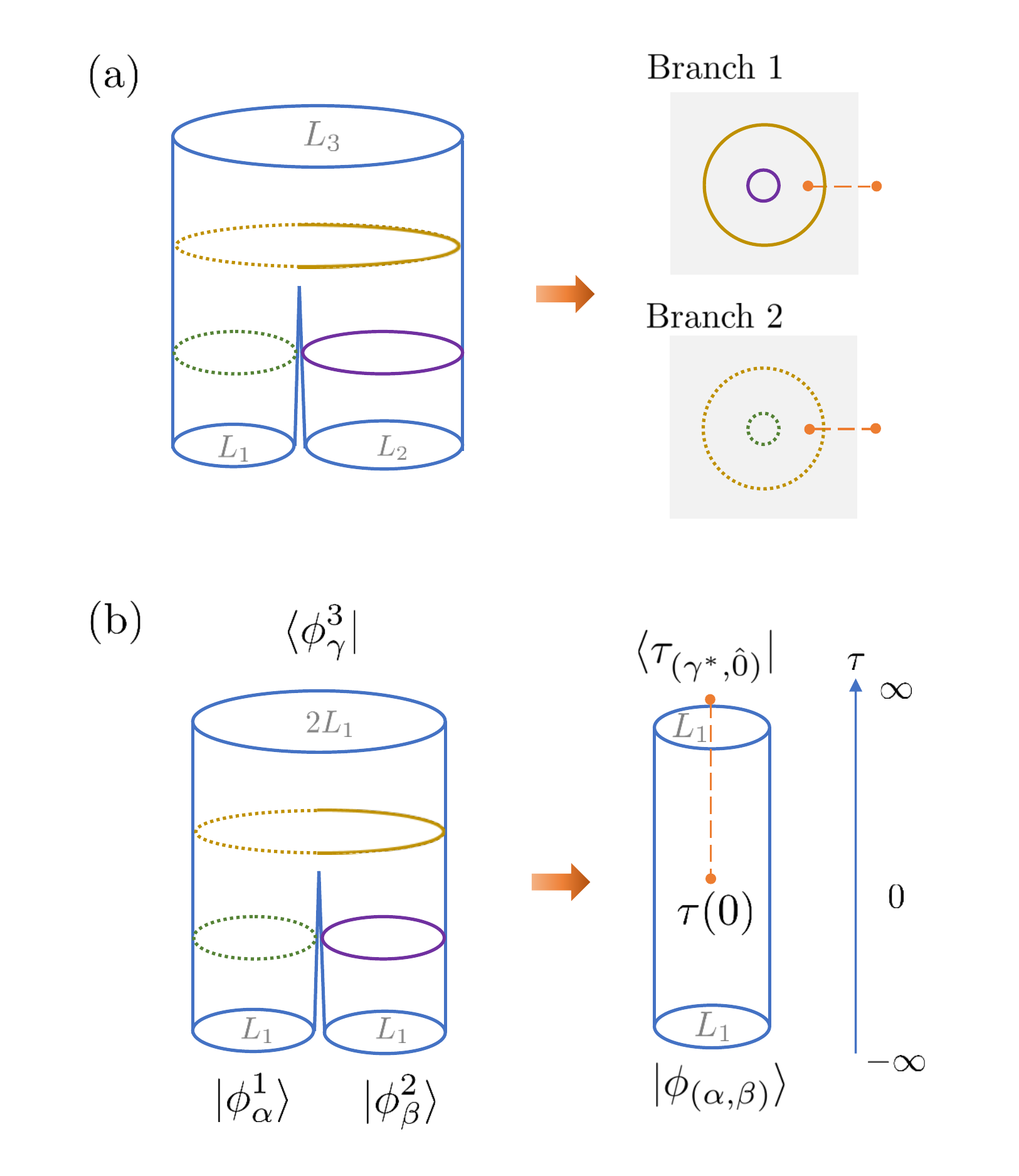}
    \caption{Mapping the path integral for the wavefunction overlap $A_{\alpha\beta\gamma}$ to the cyclic orbifold path integral. (a) The three-sided cylinder is conformally equivalent to the double-sheeted Riemann sphere with branch cut on $[1,\infty)$. The conformal transformation is achieved by $z'=e^{2\pi z/L_1}$. (b) The corresponding path integral of the cyclic orbifold for $A_{\alpha\beta\gamma}$. The insertions of branch-point twist operators at $\tau=0$ and $\tau=+\infty$ correspond to the branch cut in (a).}
    \label{fig:orbifold}
\end{figure}

\subsection{Universal finite-size corrections in wavefunction overlaps}
Now we relate the wavefunction overlaps to path integral of the cyclic orbifold. Denote the field on $\Sigma$ as 
\begin{eqnarray}
     \Phi_1(\tau,x) &=& \Phi(\tau,x),~~~ 0\leq x < L_1, \\
     \Phi_2(\tau,x) &=& \Phi(\tau,x),~~~ L_1\leq x < L_3.
\end{eqnarray}
It is then clear (as illustrated in Fig. \ref{fig:orbifold}(b)) that the fields are subject to the boundary conditions 
\begin{eqnarray}
     \Phi_i(\tau,x+L_1) &=& \Phi_i(\tau,x), ~~~ \tau< 0, \\
     \Phi_i(\tau,x+L_1) &=& \sigma\Phi_i(\tau,x), ~~~ \tau\geq 0,
\end{eqnarray}
where $\sigma\Phi_i \equiv \Phi_{i+1}$ and the subscript is understood to have periodicity $2$. Now we see the path integral on $\Sigma$ is exactly that of the cyclic orbifold on a cylinder with circumference $L_1$, with two twist operators $\tau$ inserting at $\tau=0$ and $\tau = +\infty$,
\begin{equation}
     \langle \bo^{3}|\bo^{1}\bo^{2}\rangle = \langle\bo|\tau(+\infty)\tau(0)|\bo\rangle_{\mathrm{orb}},
\end{equation}
where the twist operator changes the boundary condition.
The twist operator may contain all operators in the twisted sector,
and it can be expanded as
\begin{equation}
\label{eq:tau}
    \tau = \sum_{\alpha,\psi} a_{(\alpha,\hat{\psi})}\epsilon^{\Delta_{(\alpha,\hat{\psi})}}\tau_{(\alpha,\hat{\psi})} + \mathrm{des.}
\end{equation}
where we have introduced the UV cutoff $\epsilon$, and $a_{(\alpha,\hat{\psi})}$ is presumably non-universal constant that depends on specific lattice realizations of the CFT. Some of the coefficients may also vanish due to symmetry of the lattice model, as we later examine specific models.

Using the operator-state correspondence, Eqs.~\eqref{eq:op_state1} and \eqref{eq:op_state2}, we obtain
\begin{equation}
     \langle \bo^{3}|\bo^{1}\bo^{2}\rangle = \langle\tau_{(\bo,\hat{0})}|\tau(0)|\phi_{(\bo,\bo)_s}\rangle_{\text{orb}},
\end{equation}
where the bra is fixed to $\langle\tau_{(\bo,\hat{0})}|$ because other states in the twisted sector die off under the infinite imaginary time evolution. Using Eq.~\eqref{eq:OPEcyl} and the fact that $C_{\bo \phi \phi}=1$ for any primary operator $\phi$ (we are using $\phi=\tau_{(\bo,\hat{0})}$ here), we see
\begin{equation}
    A_{\bo\bo\bo}=\langle \bo^{3}|\bo^{1}\bo^{2}\rangle = a_{(\bo,\hat{0})}  N^{-c/8},
\end{equation}
where $N = L_1/(2\pi \epsilon)$ is the system size of the lattice model, and $c/8$ is the scaling dimension of the leading twist operator $\tau_{(\bo,\hat{0})}$. This is remarkable, since it gives a way to compute the central charge of the CFT from wavefunction overlaps.

Now we consider the general wavefunction overlap $\langle \phi^{3}_{\gamma}|\phi^{1}_{\alpha}\phi^{2}_{\beta}\rangle$. The ket state can be created by inserting $\phi_{(\alpha,\beta)}$ at $\tau=-\infty$. Hereafter we take $\phi_{(\alpha,\alpha)}$ to mean $\phi_{(\alpha,\alpha)_s}$. The bra state can be created by inserting $\phi^{*}_{(\gamma,\bo)}$ at $\tau = +\infty$\footnote{The reason that the second subindex is $\bo$ can be understood in the following way. One considers a four-sided cylinder where the third side of the three-sided cylinder is further split into two cylinders. We insert primary operator $\phi_{\alpha}$ and $\phi_{\beta}$ for the kets and $\phi^*_{\gamma}$ and $\mathbbm{1}$ for the bras. On the bra side, the two insertions fuse into $\phi^*_{\gamma}$ which gives the desired overlap $\langle \phi_{\gamma}|\phi_{\alpha}\phi_{\beta}\rangle$. 
}.
Recall that at $\tau = +\infty$ there is already a twist operator $\tau_{(\bo,\hat{0})}$ insertion, so the two operators fuse to give $\tau_{(\bo,\hat{0})} \times \phi^{*}_{(\gamma,\bo)} \ra \tau_{(\gamma^{*},\hat{0})}$+$\tau_{(\gamma^{*},\hat{1})}$, and we only keep $\tau_{(\gamma^{*},\hat{0})}$ at $\tau\ra+\infty$. Therefore, we obtain 
\begin{equation}
    \langle \phi^{3}_{\gamma}|\phi^{1}_{\alpha}\phi^{2}_{\beta}\rangle = \langle \tau_{(\gamma,\hat{0})}|\tau(0)|\phi_{(\alpha,\beta)}\rangle_{\text{orb}}.
\end{equation}
Using Eq.~\eqref{eq:tau} and \eqref{eq:OPEcyl} we obtain the expansion
\begin{equation}
    \langle \phi^{3}_{\gamma}|\phi^{1}_{\alpha}\phi^{2}_{\beta}\rangle = \sum_{\delta, \chi} a_{(\delta,\hat{\chi})} N^{-\Delta_{(\delta,\hat{\chi})}} C_{(\alpha,\beta),(\delta,\hat{\chi}),(\gamma,\hat{0})} + \cdots,
\end{equation}
where $\cdots$ stands for contributions from descendant operators.
The OPE coefficients can be computed from the OPEs in the original CFT following Ref.~\cite{dupic2018entanglement}. In particular,
\begin{equation}
    C_{(\alpha,\beta),(\bo,\hat{0}),(\gamma,\hat{0})} = 2^{-2\Delta_{\alpha}-2\Delta_{\beta}+\Delta_{\gamma}} C_{\alpha\beta\gamma},
\end{equation}
which recovers Eq.~\eqref{eq:ov3state} in the thermodynamic limit. Furthermore, we obtain the finite-size corrections of the form Eq.~\eqref{eq:FSS}, where the exponents are
\begin{equation}
\label{eq:p_set}
    p_{\alpha\beta\gamma} \in \{\Delta_{(\delta,\hat{\chi})} - \Delta_{(\bo,\hat{0})}|\, \mN_{(\alpha,\beta),(\delta,\hat{\chi}),(\gamma,\hat{0})}\neq 0\}.
\end{equation}
This is the central result of our work. We see that the exponents inside the finite-size corrections are universal, as claimed. Later we will consider various CFTs and their lattice realizations, where we numerically extract the leading exponent $p_{\alpha\beta\gamma}$ in the finite-size corrections. To compare it with Eq.~\eqref{eq:p_set}, it is useful to use the crossing symmetry,
\begin{equation}
    \mN_{(\alpha,\beta),(\delta,\hat{\chi}),(\gamma,\hat{0})} = \mN_{(\alpha,\beta),(\gamma^{*},\hat{0}),(\delta^{*},\hat{\chi})}.
\end{equation}
The leading $\Delta_{(\delta,\hat{\chi})}$ in the set of Eq.~\eqref{eq:p_set} is the lowest scaling dimension of the twist operators (except $\tau_{(\bo,\hat{0})}$) in the OPE of $\phi_{(\alpha,\beta)}$ and $\tau_{(\gamma^*,\hat{0})}$.

One simple example is $\langle \phi^{3}_{\gamma}|\bo^{1}\bo^{2}\rangle$ when $\phi_{\gamma}\neq \bo$. The fusion channel is forbidden, meaning that $\tilde{A}^{(0)}_{\bo\bo\gamma} = 0$. Since the fusion of identity operator with $\tau_{(\gamma^{*},\hat{0})}$ only gives $\tau_{(\gamma^{*},\hat{0})}$, the leading exponent in finite-size correction is
\begin{equation}
\label{eq:IIgamma}
    p_{\bo\bo\gamma} = \Delta_{(\gamma^{*},\hat{0})}- \Delta_{(\bo,\hat{0})} = \frac{\Delta_{\gamma}}{2}.
\end{equation}
This provides another consistency check of scaling dimensions of primary operators.

We stress that the result Eq.~\eqref{eq:p_set} only applies to wavefunction overlaps with $N_1=N_2$. For $N_1\neq N_2$, the exponents $p_{\alpha\beta\gamma}$ in finite-size corrections may be different. Nevertheless, the exponents are still the difference between the scaling dimensions of an operator $\tau_{(\delta,\hat{\chi})}$ in the twisted sector and $\tau_{(\bo,0)}$, while the operator $\tau_{(\delta,\hat{\chi})}$ does not necessarily satisfy the fusion constraint in Eq.~\eqref{eq:p_set}. This is examined and further explored in Appendix \ref{app:generalized}.

\subsection{Finite-size corrections to Renyi entropy of CFT}

Cyclic orbifold has also played an important role in computing the entanglement of CFT. We expect that the second Renyi entropy of CFT has similar finite-size corrections as the wavefunction overlaps, since the twist operator insertion is essentially of the same nature as Eq.~\eqref{eq:tau}. Consider, for example, the ground state of a CFT on the complex plane and the reduced density matrix on $[0,l]$,
\begin{equation}
    \rho = \tr_{\mathbb{R}-[0,l]} \, |\bo\rangle\langle\bo|.
\end{equation}
The Renyi entropies are defined by
\begin{equation}
    S_{\mathsf{N}} = \frac{1}{1-\mathsf{N}}\log \frac{\tr\,\rho^\mathsf{N}}{(\tr \rho)^{\mathsf{N}}}.
\end{equation}
Specializing 
to $\mathsf{N}=2$, the purity can be computed as the path integral of the cyclic orbifold on the complex plane with two twist operators insertions,
\begin{equation}
    \frac{\tr\,\rho^2}{(\tr \rho)^2} = \langle \bo|\tau(0)\tau(l)|\bo\rangle_{\mathrm{orb}}.
\end{equation}
Using Eq.~\eqref{eq:tau}, the correlation function can be expanded as
\begin{equation}
    \langle \bo|\tau(0)\tau(l)|\bo\rangle_{\text{orb}} = \sum_{(\alpha,\hat{\chi})} a^2_{(\alpha,\hat{\chi})}l^{-2\Delta_{(\alpha,\hat{\chi})}} + \cdots,
\end{equation}
where $\cdots$ contains contributions from descendant operators. In the limit where $l\gg\epsilon$, 
the second Renyi entropy also has universal finite-size corrections
\begin{align}
    S_2 &= \frac{c}{4} \log \frac{l}{\epsilon} - 2\log a_{(\bo,\hat{0})} 
    \nonumber \\
    &\quad 
    + \sum_{(\alpha,\hat{\chi})\neq (\bo,\hat{0})} \frac{a^2_{(\alpha,\hat{\chi})}}{a^2_{(\bo,\hat{0})}} \, l^{-2(\Delta_{(\alpha,\hat{\chi})}-\Delta_{(\bo,\hat{0})})},
\end{align}
where the last term contains the universal finite-size corrections, which has been observed in Ref.~\cite{Ohmori_2015}. As a side remark, one may consider similar finite-size corrections to other Renyi entropies on the same footing, and then analytically continue the result to the case of entanglement entropy ($\mathsf{N}=1$). Finally, we note that only the twist operators $\tau_{(\alpha,\hat{0})}$ have been considered in previous work that discusses entanglement of CFT. As we will show numerically, in the context of wavefunction overlaps the other twist operator $\tau_{(\alpha,\hat{1})}$ also plays an important role. It remains to be explored whether operators such as $\tau_{(\alpha,\hat{1})}$ can be observed in the finite-size corrections to the entanglement entropy of CFT.
\section{Applications to Ising and XXZ models}
\label{sec:Ising_main}
In this section, we apply the above general method to two specific lattice models: the Ising model and the XXZ model. 
The Ising model is the lattice realization of the $c=1/2$ Ising CFT, 
which is a minimal model. 
The XXZ model is 
a lattice realization of the $c=1$ compactified boson theory
at radius $R$, which is a rational CFT at special radii. Both the Ising model and the XXZ model at $R=2$ can be mapped to a free fermion problem. Thus we are able to obtain low-energy eigenstates 
for up to thousands of spins using the covariance matrix techniques \cite{Peschel_2009}.

We first review the operator content of each CFT and 
the $\mathsf{N}=2$ cyclic orbifold, and then show how the orbifold fusion rule determines the universal finite-size corrections in wavefunction overlaps. We then verify the leading exponent in the finite-size corrections Eq.~\eqref{eq:FSS} numerically. All the exponents agree with our analytical expression Eq.~\eqref{eq:p_set}. 

\subsection{The Ising model}
In this subsection we study the ferromagnetic Ising model. 
For later discussions, it will also be useful to study the antiferromagnetic Ising model which is described by the same CFT. 
Since the computations are largely identical, the result for the antiferromagnetic case is shown in Appendix \ref{app:antiferro}. The ferromagnetic Ising spin chain Hamiltonian at the critical point is:
\begin{equation}
    H = -\sum_{i=1}^N X_i X_{i+1}-\sum_{i=1}^N Z_i,
\end{equation}
where $X,Z$ are Pauli matrices. There are three primary operators in the Ising CFT: $\mathbbm{1},\sigma,\epsilon$, with scaling dimension $\Delta=0,\frac{1}{8},1$, respectively. 

The non-trivial OPE coefficients in the Ising CFT are:
\begin{equation}
    \begin{aligned}
     C_{\bo\bo\bo}=1,\quad C_{\epsilon\epsilon \bo}=1,\quad
        C_{\sigma\sigma \bo}=1,\quad C_{\sigma\sigma\epsilon}=\frac{1}{2},
    \end{aligned}
\end{equation}
and the modular matrices are (in the basis $\bo,\sigma,\epsilon$):
\begin{equation}
\begin{aligned}
    \mT &= \left(
    \begin{array}{ccc}
        e^{-\frac{i\pi}{24}} & 0 & 0 \\
        0 & e^{\frac{i\pi}{12}} & 0\\
        0 & 0 & e^{i\pi\frac{ 23}{24}}
    \end{array}
    \right),
    \\
    \mS &= \frac{1}{2}\left(
    \begin{array}{ccc}
        1 & 1 & \sqrt{2}  \\
        1 & 1 & -\sqrt{2}\\
        \sqrt{2} & -\sqrt{2} & 0
    \end{array}
    \right).
\end{aligned}
\end{equation}
In the following, we consider the cyclic orbifold theory of the Ising CFT. It is straightforward to compute scaling dimensions of all primary operators and the fusion rules using general framework presented in Sec.~\ref{sec:cyclic_op}. For the Ising CFT, however, it is possible to map the cyclic orbifold to a more familiar CFT, that is, a compactified boson $\mathbb{Z}_2$ orbifold at radius $R=4$. The $U(1)$ charge conservation symmetry in the latter CFT makes it more intuitive to understand the fusion rules. We have checked that the fusion rules given by Eqs.~\eqref{eqn:fusion_orbifold_1} and \eqref{eqn:fusion_orbifold_2} coincide with the those of the free boson orbifold.

\subsubsection{Orbifold operator content}
 Following the general discussion in Sec.\ \ref{sec:cyclic_op}, there exist 15 primary operators in the cyclic orbifold CFT, 9 in the untwisted sector and 6 in the twisted sector. The 15 primary operators are listed in the first column of Table.\ \ref{tbl:orb}, along with their characters $\chi$ in the last column.

\paragraph{Untwisted sector} 
The three primaries for $\alpha\neq \beta$ are
\begin{equation}
    (\bo,\sigma), \quad (\bo,\epsilon),\quad (\sigma,\epsilon),
\end{equation}
with scaling dimensions $\Delta=\frac{1}{8},1,\frac{9}{8}$. 
Additionally, there are 6 primary operators,
\begin{equation}
    (\bo,\bo)_s,\quad (\sigma,\sigma)_s,\quad (\epsilon,\epsilon)_s,
\end{equation}
with $\Delta=0,\frac{1}{4},2$ and 
\begin{equation}
    (\bo,\bo)_a,\quad (\sigma,\sigma)_a,\quad (\epsilon,\epsilon)_a,
\end{equation}
with $\Delta=4,\frac{9}{4},4$. 

\paragraph{Twisted sector} 
In the twisted sector, twist operators $\tau_{(\alpha,\hat{0})}$ have scaling dimension $\Delta=\frac{c}{8}+\frac{\Delta_\alpha}{2}$, which gives $\Delta=\frac{1}{16},\frac{1}{8},\frac{9}{16}$.
$\tau_{(\alpha,\hat{1})}$ have conformal dimension $\Delta=\frac{c}{8}+\frac{\Delta_\alpha}{2}+1$, except for $\tau_{(\bo,\hat{1})}$, which has $\Delta=\frac{c}{8}+\frac{\Delta_i}{2}+3$. 
This stems from the fact that $\bo$ corresponds to conformal vacuum $|0\rangle$ and $L_{-1}|0\rangle=0$. 
Thus, we have $\Delta=\frac{49}{16},\frac{9}{8},\frac{25}{16}$ for the primary operators $\tau_{(\alpha,\hat{1})}$. 

\subsubsection{Mapping to compactified boson orbifold}

It has been noted in Ref.~\cite{klemm1990orbifolds} that the scaling dimensions of the Ising orbifold listed above are identical to those of the compactified boson $\mathbb{Z}_2$ orbifold at radius $R=4$. Indeed the two theories are dual to each other. The free boson action is:
\begin{equation}
    S = \frac{1}{4\pi}\int d^2 z \partial \varphi \bar{\partial}\varphi.
\end{equation}
The compactified boson with radius $R$ is obtained by identifying $\varphi\equiv \varphi+2\pi n R, n\in\mathbb{Z}$. It has a $\mathbb{Z}_2$ reflection symmetry: $\varphi\ra -\varphi$, and a $U(1)$ symmetry $\varphi\ra\varphi+\alpha$. 
After gauging the $\mathbb{Z}_2$ reflection symmetry, we obtain the compactified boson orbifold theory.

The compactified boson orbifold at radius $R=\sqrt{2k}$ with $k\in\mathbb{Z}_+$ is a rational CFT with $k+7$ extended primary fields, which consists of the following five families \cite{francesco2012conformal}:
\begin{enumerate}
    \item $\phi_\lambda,\lambda=1,2,\cdots,k-1$ with dimension $\Delta_\lambda=\frac{\lambda^2}{2k}$. 
    \item $\Phi^{(1)}$ and $\Phi^{(2)}$, with dimension $\Delta=\frac{k^2}{2k}=\frac{k}{2}$.  
    \item  $\sigma^{(1)},\sigma^{(2)}$ with  dimension $\Delta=\frac{1}{8}$, and $\tau^{(1)},\tau^{(2)}$ with dimension $\Delta=\frac{9}{8}$.
    \item Identity operator $\bo$ with dimension 0.
    \item Operator $\Theta$ with  dimension $\Delta=2$. 
\end{enumerate}
When taking $k=8$, i.e., at radius $R=\sqrt{2k}=4$, the conformal dimensions of these 15 extended primaries match those of 
the 15 primaries in the cyclic Ising orbifold, as shown in Table \ref{tbl:orb}. This correspondence of conformal dimensions 
was first noticed in \cite{klemm1990orbifolds}. We took one step further and also verified numerically that this correspondence holds at the level of 
the characters $\chi$ and 
the modular matrices $\mathcal{T},\mathcal{S}$.

Mapping to the compactified boson orbifold at radius $R=4$ is useful in the problem concerned, because the fusion rule in the compactified boson orbifold is well understood, which we will now exploit to derive the universal finite-size corrections. For example, the $U(1)$ charge conservation gives:
\begin{equation}
    [\phi_\lambda]\times[\phi_\mu]=[\phi_{\lambda+\mu}]+[\phi_{\lambda-\mu}],\quad \mu\neq\lambda,2k-\lambda,
\end{equation}
and
\begin{eqnarray}
     [\Theta]\times[\phi_i]=[\phi_i].
\end{eqnarray}
The latter corresponds to the fact that $\Theta$ is charge neutral.

\begin{table}
\centering
\begin{tabular}{cccc}
\hline
\textbf{Untwisted} &&&\\
\hline
\makecell{primary\\(Ising orb.)} & $\Delta$ & \makecell{primary\\(boson orb.)} & $\chi$ \\
\hline
$(\bo,\sigma)$ & $\frac{1}{8}$ & $\sigma^{(1)}$ & $\chi_{\bo}(\tau)\chi_\sigma(\tau)$ \\
$(\bo,\epsilon)$ & $1$& $\phi_4$ & $\chi_{\bo}(\tau)\chi_\epsilon(\tau)$ \\
$(\sigma,\epsilon)$ & $\frac{9}{8}$ & $\tau^{(1)}$ & $\chi_{\sigma}(\tau)\chi_\epsilon(\tau)$ \\
$(\bo,\bo)_s$ & $0$ & $\bo$ & $\frac{1}{2}\chi_\bo^2(\tau)+\frac{1}{2}\chi_\bo(2\tau)$ \\
$(\bo,\bo)_a$ & $4$ & $\Phi^{(1)}$ & $\frac{1}{2}\chi_\bo^2(\tau)-\frac{1}{2}\chi_\bo(2\tau)$ \\
$(\sigma, \sigma)_s$ & $\frac{1}{4}$ & $\phi_2$ & $\frac{1}{2}\chi_\sigma^2(\tau)+\frac{1}{2}\chi_\sigma(2\tau)$ \\
$(\sigma, \sigma)_a$ & $\frac{9}{4}$ & $\phi_6$ & $\frac{1}{2}\chi_\sigma^2(\tau)-\frac{1}{2}\chi_\sigma(2\tau)$ \\
$(\epsilon, \epsilon)_s$ & $2$ & $\Theta$ & $\frac{1}{2}\chi_\epsilon^2(\tau)+\frac{1}{2}\chi_\epsilon(2\tau)$ \\
$(\epsilon, \epsilon)_a$ & $4$ & $\Phi^{(2)}$ & $\frac{1}{2}\chi_\epsilon^2(\tau)-\frac{1}{2}\chi_\epsilon(2\tau)$ \\
\hline
\textbf{Twisted} &&&\\
\hline
\makecell{primary\\(Ising orb.)} & $\Delta$ & \makecell{primary\\(boson orb.)} & $\chi$\\
\hline
$\tau_{(\bo,\hat{0})}$ & $\frac{1}{16}$ & $\phi_1$ & $\frac{1}{2}[\chi_\bo(\frac{\tau}{2})+T_\bo^{-\frac{1}{2}}\chi_\bo(\frac{\tau}{2}+\frac{1}{2})]$\\
$\tau_{(\bo,\hat{1})}$ & $\frac{49}{16}$ & $\phi_7$ &  $\frac{1}{2}[\chi_\bo(\frac{\tau}{2})-T_\bo^{-\frac{1}{2}}\chi_\bo(\frac{\tau}{2}+\frac{1}{2})]$\\
 $\tau_{(\sigma,\hat{0})}$ & $\frac{1}{8}$ & $\sigma^{(2)}$ & $\frac{1}{2}[\chi_\sigma(\frac{\tau}{2})+T_\sigma^{-\frac{1}{2}}\chi_\sigma(\frac{\tau}{2}+\frac{1}{2})]$\\
 $\tau_{(\sigma,\hat{1})}$ & $\frac{9}{8}$ & $\tau^{(2)}$ & $\frac{1}{2}[\chi_\sigma(\frac{\tau}{2})-T_\sigma^{-\frac{1}{2}}\chi_\sigma(\frac{\tau}{2}+\frac{1}{2})]$\\
 $\tau_{(\epsilon,\hat{0})}$ & $\frac{9}{16}$ & $\phi_3$ & $\frac{1}{2}[\chi_\epsilon(\frac{\tau}{2})+T_\epsilon^{-\frac{1}{2}}\chi_\epsilon(\frac{\tau}{2}+\frac{1}{2})]$\\
  $\tau_{(\epsilon,\hat{1})}$ & $\frac{25}{16}$ & $\phi_5$ & $\frac{1}{2}[\chi_\epsilon(\frac{\tau}{2})-T_\epsilon^{-\frac{1}{2}}\chi_\epsilon(\frac{\tau}{2}+\frac{1}{2})]$\\
\hline
\end{tabular}
\caption{Correspondence between the primary fields in the Ising cyclic $\mathbb{Z}_2$ orbifold (first column) and the free boson $\mathbb{Z}_2$ orbifold (third column) at $R=4$. $T_i=e^{2\pi i(h_i-\frac{c}{24})}$ is the  $i$-th diagonal element of the modular $\mT$ matrix.}
\label{tbl:orb}
\end{table}

\subsubsection{Universal finite-size correction}
Now we examine the universal finite-size corrections in Eq.~\eqref{eq:FSS}. As the Ising model has a $\mathbbm{Z}_2$ global symmetry, the expansion Eq.~\eqref{eq:tau} does not contain $\tau_{(\sigma,\hat{\chi})},\chi=0,1$. We further consider several examples.

As a first example, let us 
compute the universal finite-size correction to 
the wavefunction overlap $A_{\sigma\epsilon\sigma}$ for the following allowed 
OPE
channel:
\begin{equation}
    \sigma\times \epsilon\ra \sigma.
\end{equation}
To find all the scaling terms in the wavefunction overlap, we would need to find all operators $\tau_{(\alpha,\hat{\chi})}$ such that $C_{(\sigma,\epsilon),(\alpha,\hat{\chi}),(\sigma,\hat{0})}$ is non-zero. Using the crossing symmetry, we can compute $C_{(\sigma,\epsilon),(\sigma,\hat{0}),(\alpha,\hat{\chi})}$ equivalently. By mapping to compactified boson orbifold, $(\sigma,\epsilon)$ is mapped to $\tau^{(1)}$ and $\tau_{(\sigma,\hat{0})}$ is mapped to $\sigma^{(2)}$. The fusion rule reads:
\begin{equation}
    [\tau^{(1)}]\times [\sigma^{(2)}] = [\phi_1]+[\phi_3]+[\phi_5]+[\phi_7],
\end{equation}
which maps back to:
\begin{equation}
    [(\sigma,\epsilon)]\times[\tau_{(\sigma,\hat{0})}]=[\tau_{(\bo,\hat{0})}]+[\tau_{(\epsilon,\hat{0})}]+[\tau_{(\epsilon,\hat{1})}]+[\tau_{(\bo,\hat{1})}].
\end{equation}

\begin{figure}
    \centering
    \includegraphics[width=0.9\linewidth]{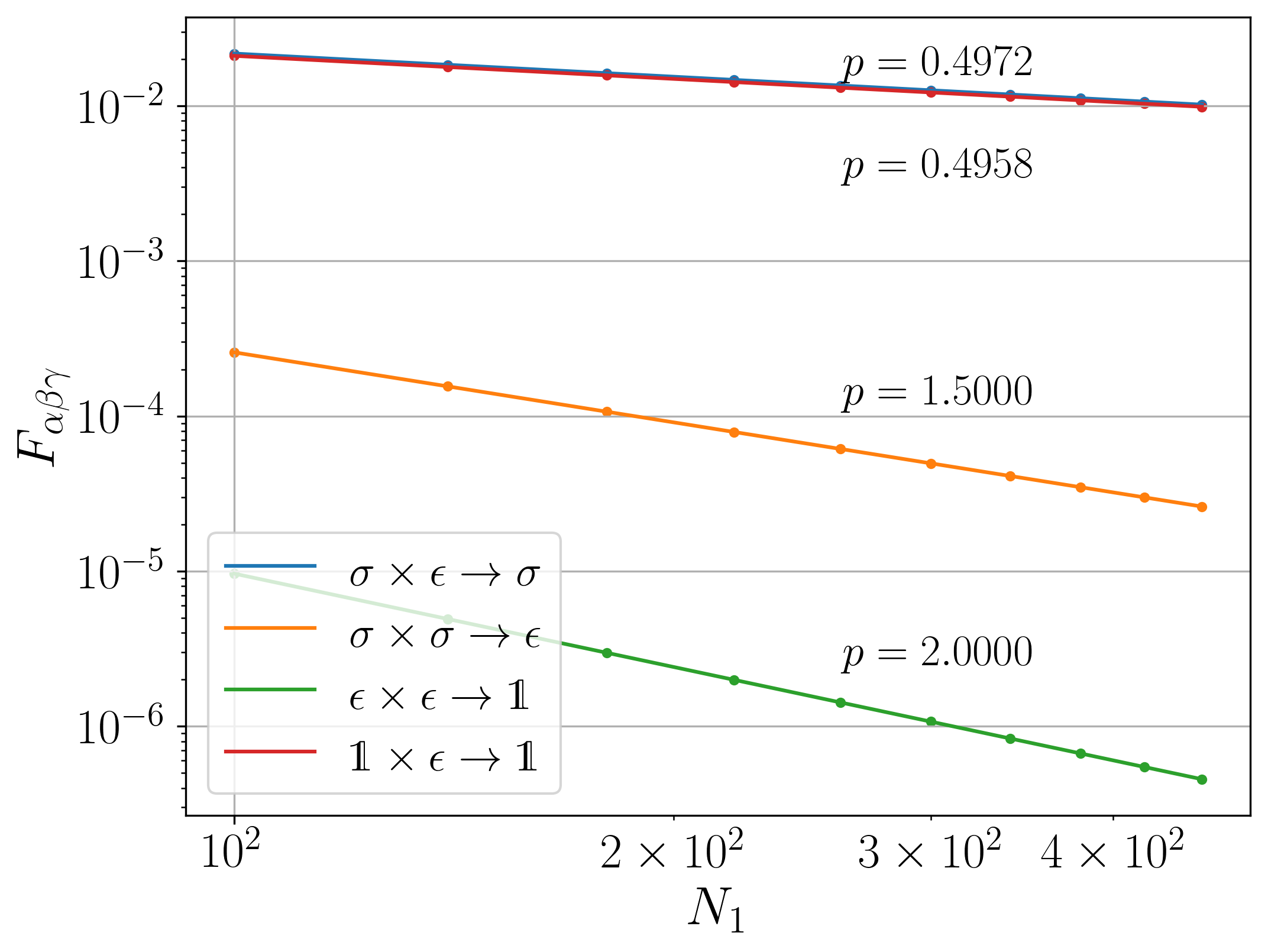}
    \caption{Finite-size corrections of wavefunction overlap in the critical Ising model, for the four channels: $\sigma\times\epsilon\ra\sigma$, $\sigma\times\sigma\ra \epsilon, \epsilon\times\epsilon\ra \bo, \bo\times\epsilon\ra \bo$ under discussion. 
    The powers of leading finite-size corrections are $p=\frac{1}{2},\frac{3}{2},2,\frac{1}{2}$, respectively, from the orbifold CFT. In the numerical computation we choose the system size $N_1\in[100,500)$, and we see the discrepancy between the theory and numerical value is less than $1\%$. 
    }
    \label{fig:Isingfsc}
\end{figure}

The four resulting primaries on the r.h.s. have scaling dimensions $\Delta=\frac{1}{16},\frac{9}{16},\frac{25}{16},\frac{49}{16}$, respectively, and the scaling of the wavefunction overlap is thus:
\begin{equation}
\begin{aligned}
    A_{\sigma\epsilon\sigma}&\approx
    a_{(\bo,\hat{0})}C_{(\sigma,\epsilon),(\bo,\hat{0}),(\sigma,\hat{0})}N^{-\frac{1}{16}}
    \\
    &\quad
    +a_{(\epsilon,\hat{0})}C_{(\sigma,\epsilon),(\epsilon,\hat{0}),(\sigma,\hat{0})}N^{-\frac{9}{16}},
\end{aligned}
\end{equation}
where we only keep 
the leading and first subleading terms. 
The leading term scales as $N^{-\frac{1}{16}}$ with power agrees with $-\frac{c}{8}$. This aligns with the fact that the wavefunction overlap for an allowed fusion process has a leading term that scales as $N^{-\frac{c}{8}}$. 
After dividing by $A_{\bo\bo\bo}$ and subtracting the leading term, the universal finite-size correction scales as $N^{-\frac{1}{2}}$, namely, $p_{\sigma\epsilon\sigma}=\frac{9}{16}-\frac{1}{16}=\frac{1}{2}$. This is exactly what we get from numerics in Fig.\ \ref{fig:Isingfsc}.

As a second example, let us study the channel:
\begin{eqnarray}
    \sigma\times \sigma\ra \epsilon,
\end{eqnarray}
where we need to find non-zero $C_{(\sigma,\sigma)_s,\tau_p,\tau_{(\epsilon,\hat{0})}}$. The fusion rule $[\phi_2]\times[\phi_3]\ra[\phi_1]+[\phi_5]$ is mapped to $[(\sigma,\sigma)_s]\times [\tau_{(\epsilon,\hat{0})}]\ra [\tau_{(\bo,\hat{0})}]+[\tau_{(\epsilon,\hat{1})}]$. This shows $\tau_p$ can take $\tau_{(\bo,\hat{0})}$ or $\tau_{(\epsilon,\hat{1})}$, and the difference of their scaling dimension predicts $p_{\sigma\sigma\epsilon}=\frac{25}{16}-\frac{1}{16}=\frac{3}{2}$.

Following the same reasoning, for the 
OPE
channel:
\begin{eqnarray}
    \epsilon\times \epsilon\ra \bo,
\end{eqnarray}
the corresponding orbifold fusion channel is
$[\Theta]\times[\phi_1]\ra[\phi_1]$, which translates into $[(\epsilon,\epsilon)_s]\times[\tau_{(\bo,\hat{0})}]\ra [\tau_{(\bo,\hat{0})}]$. This indicates $\tau_p$ lies within the conformal tower of $[\tau_{(\bo,\hat{0})}]$. Indeed, the leading exponent is controlled by a level-two descendant, which gives $p_{\epsilon\epsilon\bo}=2$. 

Finally, we consider the following forbidden channel:
\begin{equation}
    \bo\times \epsilon\ra \bo.
\end{equation}
The fusion rule $[\phi_4]\times[\phi_1]=[\phi_3]+[\phi_5]$ is translated to $[(\bo,\epsilon)]\times[\tau_{(\bo,\hat{0})}]=[\tau_{(\epsilon,\hat{0})}]+[\tau_{(\epsilon,\hat{1})}]$, which predicts the leading term in this forbidden channel scales as $N^{-\frac{9}{16}}$, i.e., $p_{\bo \epsilon \bo}=\frac{9}{16}-\frac{1}{16}=\frac{1}{2}$.

We show in Fig.\ \ref{fig:Isingfsc} the numerically obtained finite-size scaling $F_{\alpha\beta\gamma}$ defined by:
\begin{equation}
\begin{aligned}
    F_{\alpha\beta\gamma}&\equiv \frac{\langle \phi^{3}_{\gamma}|\phi^{1}_{\alpha}\phi^{2}_{\beta}\rangle}{\langle \bo^{3}|\bo^{1}\bo^{2}\rangle} - \tilde{A}^{(0)}_{\alpha\beta\gamma} \\
    & = \frac{A_{\alpha\beta\gamma}}{A_{\bo\bo\bo}}-2^{-2\Delta_{\alpha}-2\Delta_{\beta}+\Delta_{\gamma}} C_{\alpha\beta\gamma},
\end{aligned}
\end{equation}
which agrees exactly with our prediction that for the leading order,
\begin{eqnarray}
     \log F_{\alpha\beta\gamma}=-p_{\alpha\beta\gamma}\log N+\text{const.} 
\end{eqnarray}
For readers' reference, the finite-size corrections for all fusion channels are listed in Table \ref{tab:Ising-finite}.

\begin{table*}
    \centering
    \begin{tabular}{c|c|c|c}
    \hline
        overlap & \makecell{Responsible \\ operator $\tau_p$}& \makecell{Responsible\\ fusion channel} 
        & \makecell{FS correction\\ leading power $p$}\\
        \hline
        $\bo \times \bo\ra \epsilon$ & $[\tau_{(\epsilon,\hat{0})}]$ & $[(\bo,\bo)_s]\times[\tau_{(\epsilon,\hat{0})}]\ra [\tau_{(\epsilon,\hat{0})}]$ & $\frac{1}{2}$
        \\
        $\bo \times \epsilon\ra \bo$ & $[\tau_{(\epsilon,\hat{0})}]$& $[(\bo,\epsilon)]\times  [\tau_{(\epsilon,\hat{0})}]\ra [\tau_{(\bo,\hat{0})}]$& $\frac{1}{2}$
        \\
        $\epsilon\times \epsilon\ra \bo$ & $[\tau_{(\bo,\hat{0})}]$ des. & $[(\epsilon,\epsilon)_s]\times[\tau_{(\bo,\hat{0})}]\ra [\tau_{(\bo,\hat{0})}] $ & $2$
        \\
        $\epsilon\times \epsilon\ra \epsilon$ & $[\tau_{(\epsilon,\hat{0})}]$& $[(\epsilon,\epsilon)_s]\times[\tau_{(\epsilon,\hat{0})}]\ra [\tau_{(\epsilon,\hat{0})}]$ & $\frac{1}{2}$
        \\
        $\bo\times \epsilon\ra \epsilon$ & $[\tau_{(\bo,\hat{1})}]$ & $[(\bo,\epsilon)]\times[\tau_{(\bo,\hat{1})}]\ra[\tau_{(\epsilon,\hat{0})}] $ & $3$
        \\
        $\bo\times \sigma\ra\sigma$ & $[\tau_{(\epsilon,\hat{0})}]$ & $[(\bo,\sigma)]\times[\tau_{(\epsilon,\hat{0})}]\ra[\tau_{(\sigma,\hat{0})}]$ & $\frac{1}{2}$
        \\
        $\sigma\times \epsilon\ra \sigma$  & $[\tau_{(\epsilon,\hat{0})}]$ & $[(\sigma,\epsilon)]\times[\tau_{(\epsilon,\hat{0})}]\ra[\tau_{(\sigma,\hat{0})}]$ & $\frac{1}{2}$
        \\
        $\sigma\times \sigma\ra \bo$ & $[\tau_{(\epsilon,\hat{0})}]$ & $[(\sigma,\sigma)_s]\times[\tau_{(\epsilon,\hat{0})}] \ra [\tau_{(\bo,\hat{0})}]$ & $\frac{1}{2}$
        \\
        $\sigma\times \sigma\ra\epsilon$ & $[\tau_{(\epsilon,\hat{1})}]$ & $[(\sigma,\sigma)_s]\times [\tau_{(\epsilon,\hat{1})}]\ra [\tau_{(\epsilon,\hat{0})}] $ & $\frac{3}{2}$
        \\ \hline
    \end{tabular}
    \caption{Universal finite-size corrections to wavefunction overlaps of the ciritical Ising model for all fusion channels. The wavefunction overlaps for $\bo\times\bo\ra\sigma$ 
    and $\sigma\times\sigma\ra\sigma$ are strictly zero due to the global $\mathbb{Z}_2$ symmetry in the Ising model.}
    \label{tab:Ising-finite}
\end{table*}

\subsection{The XXZ model}

The XXZ model is a lattice realization of $c=1$ compactified boson with radius $R$. The free boson has a $U(1)$ current $J=i\partial \varphi$. In the spin basis, the Hamiltonian is:
\begin{equation}
    H = -\sum_{i=1}^N (X_i X_{i+1}+Y_i Y_{i+1}+\Delta Z_i Z_{i+1}),
\end{equation}
where the radius $R$ is related to $\Delta$ via $\Delta = \cos(2\pi/R^2)$. At $\Delta=0$ (namely, $k=2, R=\sqrt{2k}=2$), the spin Hamiltonian can be Jordan-Wigner transformed into the free fermion Hamiltonian:
\begin{equation}
    H = -2\sum_{i=1}^N (c_i^\dg c_{i+1}+c_{i+1}^\dg c_i).
\end{equation}
In the following, we will focus on this free fermion point for its numerical efficiency.

\subsubsection{Operator content}
\paragraph{Original theory} At radius $R=\sqrt{2k}$, the valid chiral and anti-chiral momentum modes in the compactified boson theory are:
\begin{equation}
\begin{aligned}
    &p=\frac{m}{R}+\frac{nR}{2}=\frac{m}{2}+n,\\
    &\bar{p}=\frac{m}{R}-\frac{nR}{2}=\frac{m}{2}-n,\quad m,n\in\mathbb{Z},
\end{aligned}
\end{equation}
where $m$ is the quantized momentum and $n$ is the quantized winding. The operators can be labeled by $\mathcal{O}_{(m,n)}$.
The partition function is obtained after summing over contribution from valid momentum modes:
\begin{eqnarray}
    Z=\frac{1}{|\eta|^2}\sum_{m,n}q^{\frac{1}{2}(\frac{m}{R}+\frac{nR}{2})^2}\bar{q}^{\frac{1}{2}(\frac{m}{R}-\frac{nR}{2})^2},
\end{eqnarray}
where $\eta(\tau)$ is the Dedekind eta function.
Equivalently, for integer $k$, this $U(1)$ chiral algebra can be viewed as being extended by two vertex operators $\lbrace e^{i\sqrt{2k}\varphi},e^{-i\sqrt{2k}\bar{\varphi}}\rbrace$, which allows us to interpret it as $u(1)_k$ rational CFT and block diagonalize the partition function:
\begin{eqnarray}
    Z=\frac{1}{|\eta|^2}\sum_{l=0}^{2k-1} |\Theta_{l,k}|^2,\quad
    \Theta_{l,k}=\sum_{n} q^{k(n+\frac{l}{2k})^2}.
\end{eqnarray}
From the block-diagonal partition function, we read off $2k$ extended primaries. For the $u(1)_2$ rational CFT at $k=2$, there are $2k=4$ extended primaries with scaling dimension $\Delta_l=\frac{l^2}{2k}=\frac{l^2}{4}, l=0,1,2,3$. The first several modes in each extended primaries are:
\begin{equation}
    \begin{tabular}{c|l}
    \hline
         &  $(m,n)$\\
         \hline
        $\bo$ & (0,0), (0,2), (2,1), (4,0)\\
        $\mathcal{O}_1$ & (1,0), $(-3,0)$\\
        $\mathcal{O}_2$ & (2,0), $(-2,0)$, (0,1)\\
        $\mathcal{O}_3$ & (3,0), $(-1,0)$, (1,1)\\
        \hline
    \end{tabular}
    \label{eqn:xxztbl}
\end{equation}
Namely, $\mathcal{O}_{(1,0)},\mathcal{O}_{(-3,0)}$ belong to the conformal family $\mathcal{O}_1$, etc. 

The modular matrix $\mT$ can be obtained directly from the scaling dimension: $\mT_{ll}=\exp{(2\pi i (\Delta_l/2-c/24))}$, 
and the modular matrix $\mS$ is:
\begin{eqnarray}
    \mS_{l,l'} = \frac{1}{\sqrt{2k}}\exp{(-i\pi\frac{ll'}{k})}. 
\end{eqnarray}
The original theory modular matrices allow us to compute the modular matrices in the orbifolded theory.

For later discussions, we focus on the wavefunction overlap associated with current operator $J=i\partial\varphi$. Specifically, we consider the process:
\begin{equation}
    J\times J\ra \bo.
\end{equation}
In the $U(1)$ Kac-Moody algebra, $J$ is the Kac-Moody descendant of $\mathcal{O}_{(0,0)}=\bo$ 
(while not being the Virasoro descendant). 

\paragraph{Orbifold theory} Following the general procedure in Sec.~\ref{sec:cyclic_op}, the untwisted sector has 14 primary operators, including operators like $(\bo,\mathcal{O}_1),(\mathcal{O}_1,\mathcal{O}_2),(\mathcal{O}_1,\mathcal{O}_1)_{s/a}$, etc.; and there are 8 primaries in the twisted sector:
\begin{equation}
    \begin{tabular}{c|c|c|c|c}
    \hline
        &$\tau_{(\bo,\hat{0})}$ &  $\tau_{(\bo,\hat{1})}$& $\tau_{(\mathcal{O}_1,\hat{0})}$& $\tau_{(\mathcal{O}_1,\hat{1})}$ \\
        \hline
        $\Delta$ & $\frac{1}{8}$ & $\frac{9}{8}$& $\frac{1}{4}$ & $\frac{5}{4}$ \\
        \hline
        & $\tau_{(\mathcal{O}_2,\hat{0})}$& $\tau_{(\mathcal{O}_2,\hat{1})}$ & $\tau_{(\mathcal{O}_3,\hat{0})}$& $\tau_{(\mathcal{O}_3,\hat{1})}$\\
        \hline
        $\Delta$& $\frac{5}{8}$ & $\frac{21}{8}$ & $\frac{5}{4}$ & $\frac{13}{4}$\\
        \hline
    \end{tabular}
\end{equation}
Note that the difference between $\Delta$ for $\tau_{(\bo,\hat{0})}$ and $\tau_{(\bo,\hat{1})}$ is $1$ rather than $3$ as in the Ising orbifold. This is because we are considering the Kac-Moody algebra and $J$ is a level-1 descendant of $\bo$ with respected to the extended symmetry. 

\subsubsection{Universal finite-size correction}

Since the XXZ model has a $U(1)$ global symmetry (which corresponds to $m$ charge conservation in primary operators $\mathcal{O}_{(m,n)}$), the expansion Eq.~\eqref{eq:tau} does not contain $\tau_{(\mathcal{O}_i,\hat{\chi})},i=1,2,3, \chi=0,1$. As a first example, consider 
the OPE
channel:
\begin{eqnarray}
    \mathcal{O}_{(1,0)}\times\mathcal{O}_{(-1,0)}\ra \bo.
\end{eqnarray}
From Eq.\ \eqref{eqn:xxztbl}, $\mathcal{O}_{(1,0)}$ belongs to the conformal family $\mathcal{O}_1$ and $\mathcal{O}_{(-1,0)}$ belongs to the conformal family $\mathcal{O}_3$. We thus need to find $\tau_p$ such that $[(\mathcal{O}_1,\mathcal{O}_3)]\times[\tau_p]\ra [\tau_{(\bo,\hat{0})}]$ is valid in the orbifold theory, namely, $C_{(\mathcal{O}_1,\mathcal{O}_3),\tau_p,(\bo,\hat{0})}$ is non-zero. Using the crossing symmetry $C_{(\mathcal{O}_1,\mathcal{O}_3),\tau_p,(\bo,\hat{0})}=C_{(\mathcal{O}_1,\mathcal{O}_3),(\bo,\hat{0}),\tau_{p^*}}$ and orbifold fusion rule:
\begin{equation}
    [(\mathcal{O}_1,\mathcal{O}_3)]\times[\tau_{(\bo,\hat{0})}]=[\tau_{(\bo,\hat{0})}]+[\tau_{(\bo,\hat{1})}],
\end{equation}
we see $\tau_p$ can be either $\tau_{(\bo,\hat{0})}$ or $\tau_{(\bo,\hat{1})}$. The scaling dimension of the leading term $\tau_{(\bo,\hat{0})}$ gives the leading scaling $\Delta = \frac{1}{8}$, which agrees with $c/8$. The difference between the scaling dimensions of $\tau_{(\bo,\hat{0})}$ and $\tau_{(\bo,\hat{1})}$ gives the finite-size correction with power $p=\frac{9}{8}-\frac{1}{8}=1$. 

We may also consider the channel involving current operator $J$, which belongs to the conformal family $\bo$. Consider the OPE
channel:
\begin{eqnarray}
    J\times \mathcal{O}_{(1,0)}\ra\mathcal{O}_{(1,0)},
\end{eqnarray}
we need to find non-zero $C_{(\bo,\mathcal{O}_1),\tau_p,(\mathcal{O}_1,\hat{0})}$, which --  by crossing symmetry -- is, $C_{(\bo,\mathcal{O}_1),(\mathcal{O}_3,\hat{0}),\tau_{p^*}}$. The relevant orbifold fusion rule is:
\begin{eqnarray}
    [(\bo,\mathcal{O}_1)]\times[\tau_{(\mathcal{O}_3,\hat{0})}]=[\tau_{(\bo,\hat{0})}]+[\tau_{(\bo,\hat{1})}].
\end{eqnarray}
Again, this gives the finite-size correction with power $p=1$ as the previous example. 

In the third example, let us consider:
\begin{eqnarray}
    J\times J \ra \bo,
\end{eqnarray}
where we need to find non-zero $C_{(\bo,\bo)_s,\tau_p,(\bo,\hat{0})}$. There is only one valid $\tau_p=[\tau_{(\bo,\hat{0})}]$ from:
\begin{equation}
    [(\bo,\bo)_s]\times[\tau_{(\bo,\hat{0})}]=[\tau_{(\bo,\hat{0})}].
\end{equation}
In this case, we observe $p=2$, which means a level-2 descendant in the expansion Eq.~\eqref{eq:tau} is responsible.
Similarly, the process $J\times\mathcal{O}_{(2,1)}\ra\mathcal{O}_{(2,1)}$ also has correction $p=2$ using the same orbifold fusion rule. 

\begin{figure}[hbt]
    \centering
    \includegraphics[width=\columnwidth]{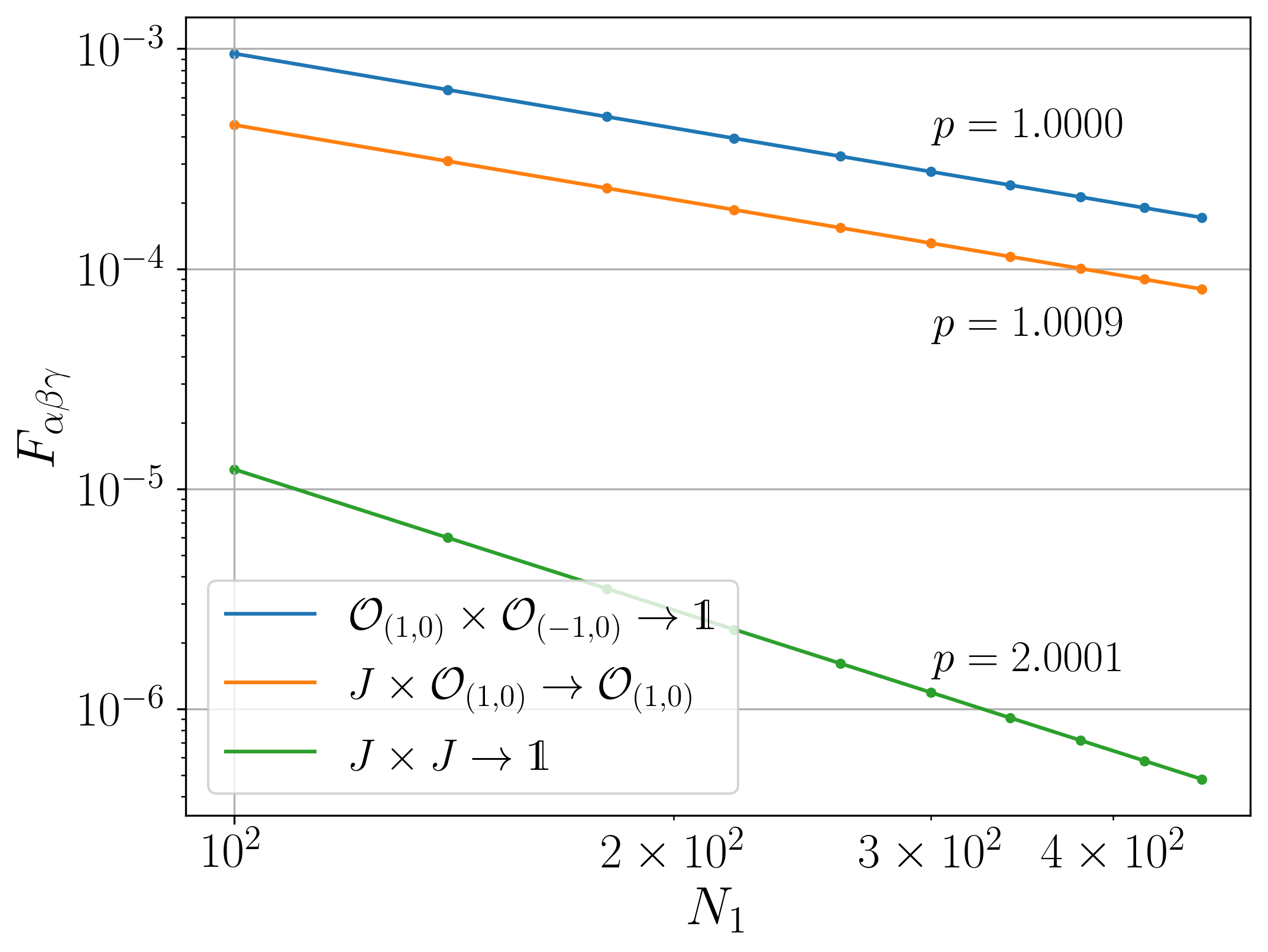}
    \caption{Finite-size correction of wavefunction overlap of the XXZ model at $\Delta=0$, for the three channels under discussion. The system size is chosen in the range $N_1\in[100,500)$. The discrepancy between the theory and numerical value is less than $0.1\%$.
    }
    \label{fig:xx_error}
\end{figure}

For other radius $R$, we have to resort to exact diagonalization or tensor network methods. We have also examined the case of $R=\sqrt{6}$ and checked that the finite-size correction exponents for various channels, in particular, the channel involving $J$, agree with analytical predictions.

\section{Applications to the Haagerup model}
\label{sec:haagerup_main}
In the previous section, we have examined the universal finite-size corrections in the Ising model and XXZ model. We now move on to the Haagerup model, which is a recently proposed \cite{huang2021numerical, vanhove2021critical} 
anyonic chain that is believed to be described by a CFT 
with $c\sim 2$. 
The input data for the lattice Hamiltonian is the $\mathcal{H}_3$ \textit{fusion category}. The fusion rule in this category is non-commutative thus does not admit braiding. 
The operator content of the Haagerup CFT is still obscure due to strong finite-size corrections as only very small systems are numerically accessible. Due to its $\mathbb{Z}_3$ global symmetry, one of the candidate theory is the $\mathbb{Z}_3$ orbifold of toroidal compactified sigma model, with central charge $c=2$. The candidate has two current operators $J$ and $J^{*}$ with scaling dimension and conformal spin $\Delta=s=1$. The holomorphic nature of the current operators strongly constrain possible fusion channels involving these operators, which enables us to check it numerically.

In what follows, we use the wavefunction overlaps to compute central charge, conformal dimensions of certain primary states. We present strong evidence that the lowest operator with $s=1$ is not a holomorphic field, thus ruling out the possibility of this $\mathbb{Z}_3$ orbifold. Nevertheless, we will denote the operators as $J$ and $J^{*}$. 
Our results give an upper bound of the scaling dimension $\Delta_{J}\leq 1.4$, and indicate that the central charge is roughly $c \approx 2.1$. Our code for numerical calculation is available in GitHub \footnote{\url{https://github.com/YuhanLiuSYSU/Haagerup_wave_function_overlap}}.


\subsection{Haagerup anyon chain and spectrum}
\subsubsection{$\mathcal{H}_3$ fusion category}
In the following we give a short review on $\mathcal{H}_3$ fusion category \cite{huang2020f, osborne2019f}.
There are six simple objects (anyons) in this category:
\begin{equation}
    1,a,a^2,\rho, a\rho, a^2\rho.
\end{equation}
We will use $a^2$ and $a^*$ interchangeably. $1,a,a^*$ are invertible objects and $\rho,a\rho,a^*\rho$ are non-invertible simple objects. Their quantum dimensions are: $d_1=d_a=d_{a^*}=1$, $d_\rho=d_{a\rho}=d_{a^*\rho}=\frac{3+\sqrt{13}}{2}$. The non-trivial fusion rules read:
\begin{equation}
    a^3=1,\quad a\rho=\rho a^*\quad \rho^2=1+(1+a+a^2)\rho.
\end{equation}
Note that for example, $a$ and $\rho$ does not commute, $a\rho\neq \rho a$. Given the fusion rule, the $F$-symbols can be determined by the pentagon equations (self-consistency requirement).

\begin{figure}
\begin{subfigure}[b]{\linewidth}
     \centering
     \includegraphics[width=\textwidth]{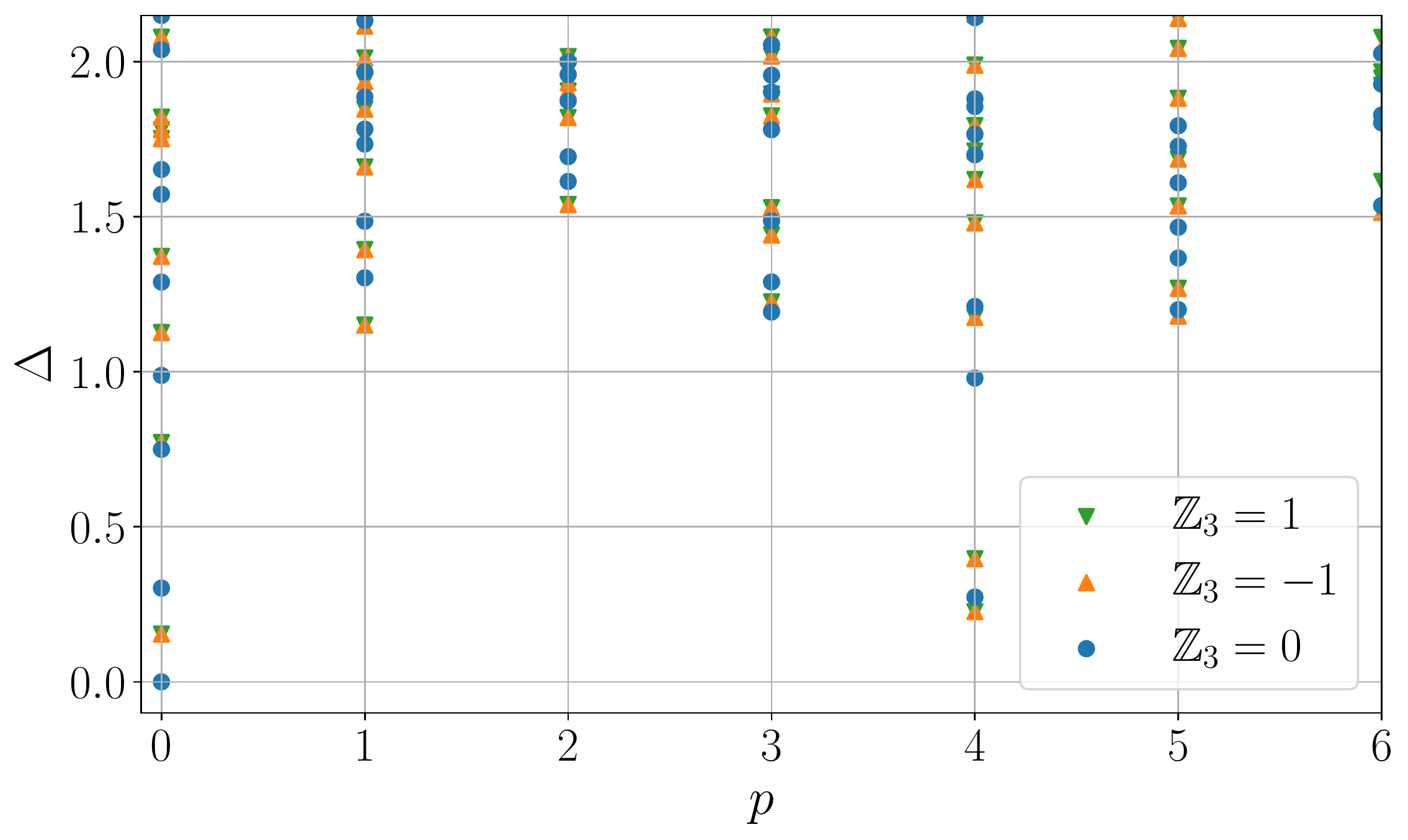}
\end{subfigure}
\begin{subfigure}[b]{\linewidth}
     \centering
     \includegraphics[width=\textwidth]{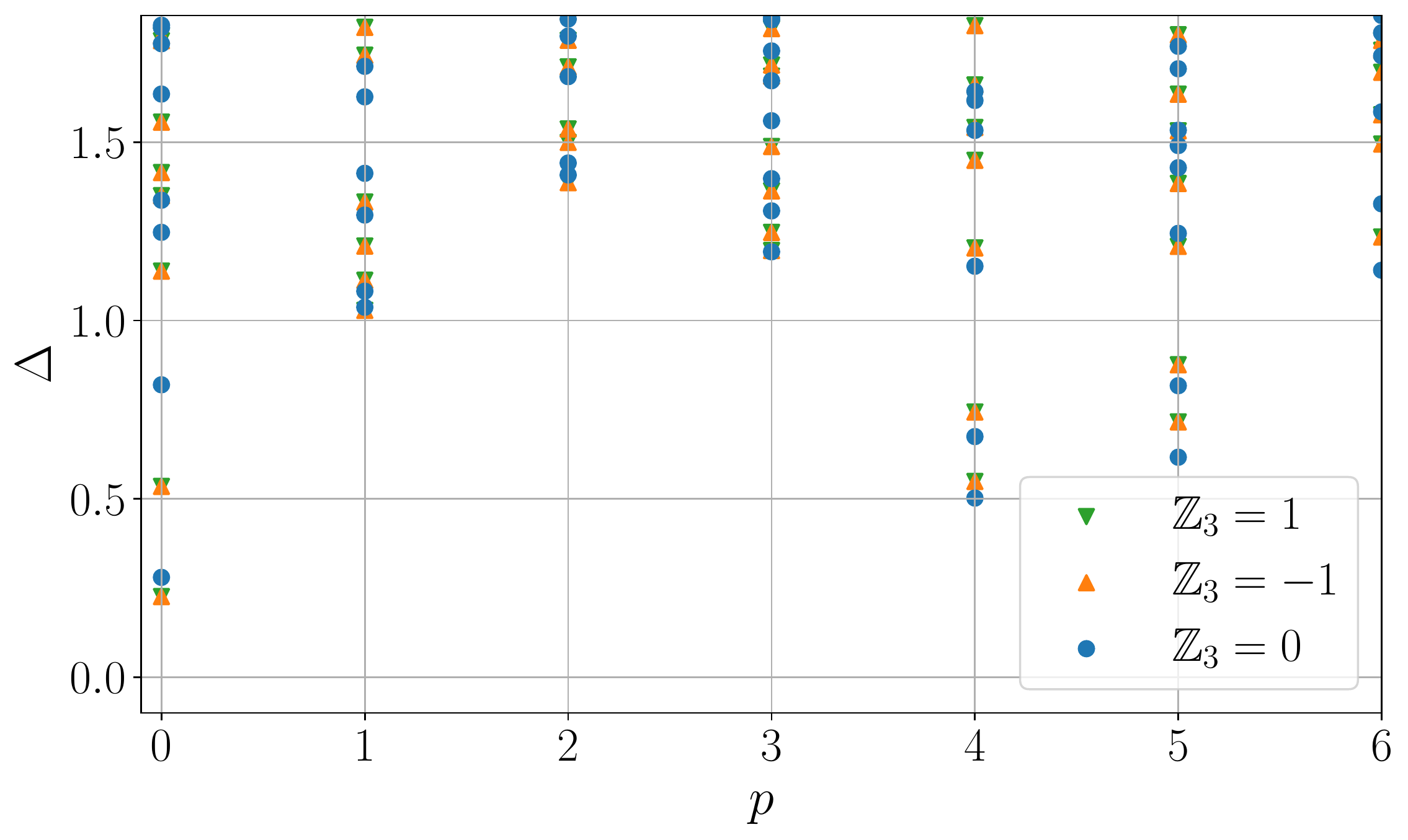}
\end{subfigure}
    \caption{Energy-momentum spectrum of the Haagerup model 
    at (a) $N=12$, and (b) $N=13$. 
     }
    \label{fig:hgr12}
\end{figure}

\subsubsection{Anyon chain construction and spectrum}

Given a fusion category as input data, the anyon chain Hamiltonian is constructed using the fusion rule and the $F$-symbol \cite{feiguin2007interacting,buican2017anyonic,huang2021numerical}. The Hilbert space consists of the anyon chains $|a_1,a_2,\cdots,a_N\rangle$ that satisfy the following fusion constraint:
\begin{eqnarray}
    \includegraphics[width=0.5\linewidth]{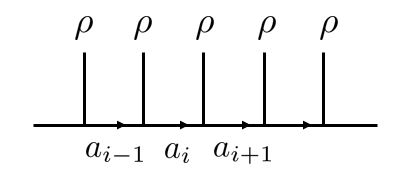}
\end{eqnarray}
Namely, the fusion $a_i\times\rho$ needs to contain $\rho$.
We assume periodic boundary condition for the anyon chain. 

The Hamiltonian is then constructed as sum of projectors acting on three subsequent anyon sites \cite{huang2021numerical,feiguin2007interacting}:
\begin{equation}
\begin{aligned}
    &H = -\sum_i P_\rho^{(i)},
    \\
    &P_c^{(i)}|a_{i-1}a_i a_{i+1}\rangle=\\
    &\quad \sum_{a_i'}[F_{a_{i+1}}^{a_{i-1}\rho\rho}]_{a_i c}[F_{a_{i+1}}^{a_{i-1}\rho\rho}]^*_{a_i' c}|a_{i-1} a_i' a_{i+1}\rangle.
\end{aligned}
\end{equation}
The projector $P_\rho$ projects the fusion $\rho\times\rho$ into channel $\rho$. The above anyonic chain construction is standard, and can be regarded as a generalization of the spin-$1/2$ Heisenberg chain \cite{buican2017anyonic}. 

It is nontrivial and has to be verified case-by-case 
whether an anyonic chain Hamiltonian corresponds to a lattice realization of a CFT. One should distinguish the input fusion category with the Moore-Seiberg tensor category of a rational CFT, where the object of the Moore-Seiberg tensor category are in one-to-one correspondence with primary operators which is not true for the input fusion category. 
For the Haagerup model, previous evidence of CFT comes from ground state energy and entanglement entropy \cite{huang2021numerical}. 
We will see in the following that the ground state wavefunction overlap $A_{\bo\bo\bo}$ provides another evidence for CFT with consistent central charge. 

By the Haagerup fusion rules, the Hamiltonian has a global $\mathbb{Z}_3$ symmetry:
\begin{eqnarray}
    \gamma:1\ra a \ra a^*\ra 1,\quad \rho\ra a\rho\ra a^*\rho\ra \rho.
\end{eqnarray}
This can also be seen at the level of $F$-symbols. There is another intrinsic shift $\mathbb{Z}_3$ symmetry $s$, that is, translation by $3n,3n+1$ and $3n+2$ sites. The shift symmetry is nontrivial because the energy spectrum depends on $N$ mod $3$. Similar phenomena occur in the antiferromagnetic Ising model, which we review in Appendix \ref{app:antiferro}. To gauge the shift $\mathbb{Z}_3$ symmetry $s$, we keep only the states near $P=0$ at system size $N=3n,3n+1,3n+2$, which correspond to the original Hilbert space $\mathcal{H}$, shift $\mathbb{Z}_3$ twisted Hilbert space $\mathcal{H}_s$ and $\mathcal{H}_{s^2}$. It is clear from the $\mathbb{Z}_3$ symmetry that $\mathcal{H}_s=\mathcal{H}_{s^2}$.

The energy-momentum spectrum of $N=12$ ($\mathcal{H}$) and $N=13$ ($\mathcal{H}_s$) anyon chain is shown in Fig.\ \ref{fig:hgr12}. To better understand the conformal towers, we also use 
the Koo-Saleur lattice Virasoro generators as in Refs.~\cite{koo1994,milsted2017,Zou_2019},
\begin{equation}
    H_n = \frac{N}{2\pi}\sum_{j=1}^N h_j e^{ijn\frac{2\pi}{N}},
\end{equation}
where 
$h_j$ is the Hamiltonian 
density operator at site $j$.
These operators
correspond 
to linear combinations of the Virasoro generators,
\begin{equation}
    H_n=L_n+\bar{L}_{-n}-\frac{c}{12}\delta_{n,0}.
\end{equation}

There are several numerical observations with the Koo-Saleur generators:
\begin{enumerate}
    \item There is only one spin-2 low-energy state that has nonzero overlap with $H_{-2}|\bo\rangle$, which we identify as the stress tensor state $|T\rangle$. This is used to normalize the energies in the plot by utilizing $|T\rangle$ has $\Delta=2$. $H_{-1}$ acting on $|\bo\rangle$ gives zero overlap for every state in the spectrum, confirming this state is the conformal vacuum. 
    
    \item In the spectrum of $N=13$ chain (shift $\mathbb{Z}_3$ twisted sector $\mathcal{H}_s$), there exists a state with $\mathbb{Z}_3=0$ being close to $P=1,\Delta=1$. This is proposed to be the chiral current operator $J$. There is a similar state in the $N=14$ chain, which is proposed to be $J^*$. 
    
    \item In the $N=13$ chain, $H_{-1}$ acting on the lowest primary state in $\mathbb{Z}_3=0,P=0$ gives the largest overlap with the second lowest state in $\mathbb{Z}_3=0, P=1$. This overlap is significantly larger than that with the first excited state with $\mathbb{Z}_3=0, P=1$. This indicates the lowest state in $\mathbb{Z}_3=0,P=1$ is not a descendant of the primary state. Thus it is a primary state.
\end{enumerate}

\subsubsection{$\mathbb{Z}_3$ orbifold of sigma model}
The $\mathbb{Z}_3$ orbifold theory referred in this paper is the nonlinear sigma model on torus target space 
modded out the $\mathbb{Z}_3$ rotational symmetry on the target space. We will briefly review the the relevant features of this CFT below.

Let us start with the toroidal compactified boson theory, namely, there are two bosonic fields and the target space is $T^2$ \cite{dulat2000crystallographic}. Recall that in the familiar example of compactified boson, the target space is $S^1$, which is equivalent to saying that the boson field $\Phi$ can only take value in $\Phi\in[0,2\pi R)$, and $\Phi\sim\Phi+2n\pi R,n\in\mathbb{Z}$ are identified. Analogously, the toroidal compactification free boson theory has two massless scalar fields $\Phi^\mu,\mu\in\lbrace 1,2\rbrace$. For the target space torus $T^2$ generated by $\bm{\lambda}_1,\bm{\lambda}_2$, we need to identify:
\begin{equation}
    \left(\begin{array}{c}
         \Phi^1 \\
         \Phi^2 
    \end{array} \right)\sim 
     \left(\begin{array}{c}
         \Phi^1 \\
         \Phi^2 
    \end{array} \right)+n_1\bm{\lambda}_1+n_2\bm{\lambda}_2,\quad n_1\in\mathbb{Z},n_2\in\mathbb{Z}.
\end{equation}
This compactification gives constraint on the momentum modes $p,\bar{p}$, from which one can write down the partition function \cite{dulat2000crystallographic}. 

When the target space torus $T^2$ is generated by basis vectors proportional to $\bm{\lambda}_1=(1,0)^T$, $\bm{\lambda}_2=(-\frac{1}{2},\frac{\sqrt{3}}{2})^T$, the theory has an additional $\mathbb{Z}_3$ rotational symmetry. Gauging (modding out) this $\mathbb{Z}_3$ rotational symmetry gives the $\mathbb{Z}_3$ orbifold theory. The partition function of this rational CFT can be written out explicitly, from which we can read out the operator content. We list the several important features that are relevant to our problem, and refer the interested readers to Ref.\ \cite{dulat2000crystallographic} for more details. The features are:
\begin{enumerate}
    \item The spectrum includes two chiral current operators: $J_1=i\partial\Phi^1$ and $J_2=i\partial \Phi^2$. This is true for all toroidal compactified theory, independent of the $\mathbb{Z}_3$ symmetry.
    \item In the $\mathbb{Z}_3$ twisted sector, the lowest energy state is three-fold degenerate, with $(h,\bar{h})=(\frac{1}{9},\frac{1}{9})$. Similar feature appears in $\mathbb{Z}_3^2$ twisted sector. The degeneracy comes from the three target space fixed points under rotation. 
\end{enumerate}

The exact degeneracy in the second feature is hard to confirm numerically, as the states suffer from strong finite-size corrections and the energies do not exactly coincide. However, the wavefunction overlap provides a way to check the first feature, with relatively small finite-size corrections. We will see below that the overlap indicates that the lowest spin-1 operator is not a current, which contradicts the proposal that Haagerup model is described by this $\mathbb{Z}_3$ orbifold. 

\subsection{Wavefunction overlaps}

We now turn to the wavefunction overlaps, from which we can extract conformal data, such as central charge, scaling dimensions and OPE coefficients. We perform exact diagonalization for small system sizes $N\leq 15$ and the periodic uniform matrix product state (puMPS) techniques for up to $N=27$. The latter is briefly reviewed in Appendix \ref{app:pumps}.

Firstly, from the conformal vacuum state overlap, we can extract the central charge via:
\begin{eqnarray}
A_{\bo\bo\bo}\propto N^{-\frac{c}{8}}.     
\end{eqnarray}
We obtain $c=2.06$, in a reasonable agreement with previous results, which give $c=2.03$ \cite{vanhove2021critical} and $c=2.11$ \cite{huang2021numerical}. 
\begin{figure}
    \centering
    \includegraphics[width = 0.9\linewidth]{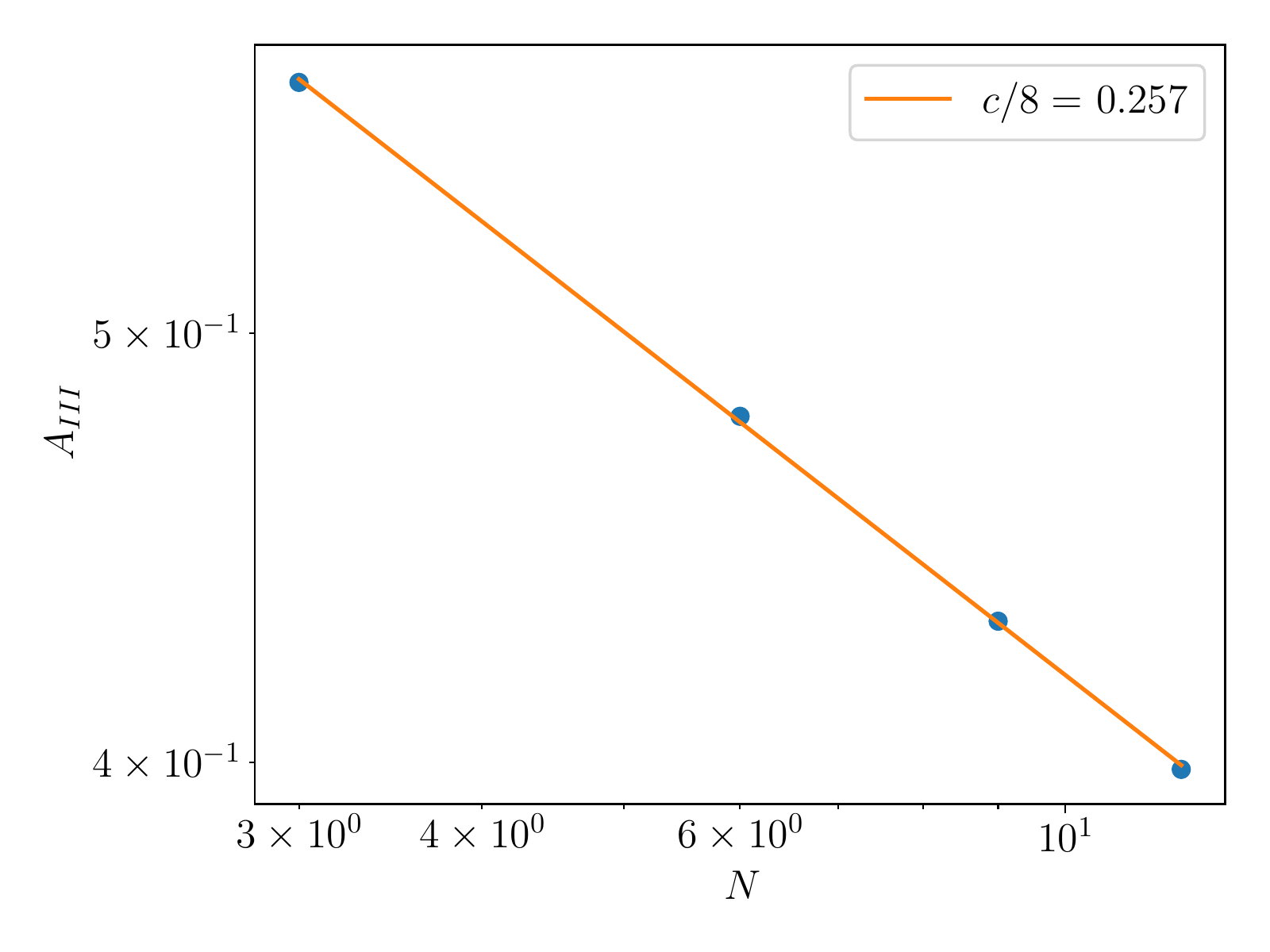}
    \caption{Ground state overlap $A_{\bo\bo\bo}$ for the Haagerup model with $N,N$ and $2N$ spins, where $N=3,6,9,12$. The first two points are obtained by exact diagonalization, and the last two points are obtained by puMPS with maximal bond dimension $D=66$.}
    \label{fig:Haagerup_AIII}
\end{figure}
If we discard the data from the smallest system size ($N=3$), then we obtain $c=2.12$ by fitting the remaining three points. Given that current numerical methods all give $c>2$, it is possible that the true central charge is not exactly at $c=2$, but $c\approx 2.1$. 

\begin{figure}
    \centering
    \includegraphics[width=0.9\linewidth]{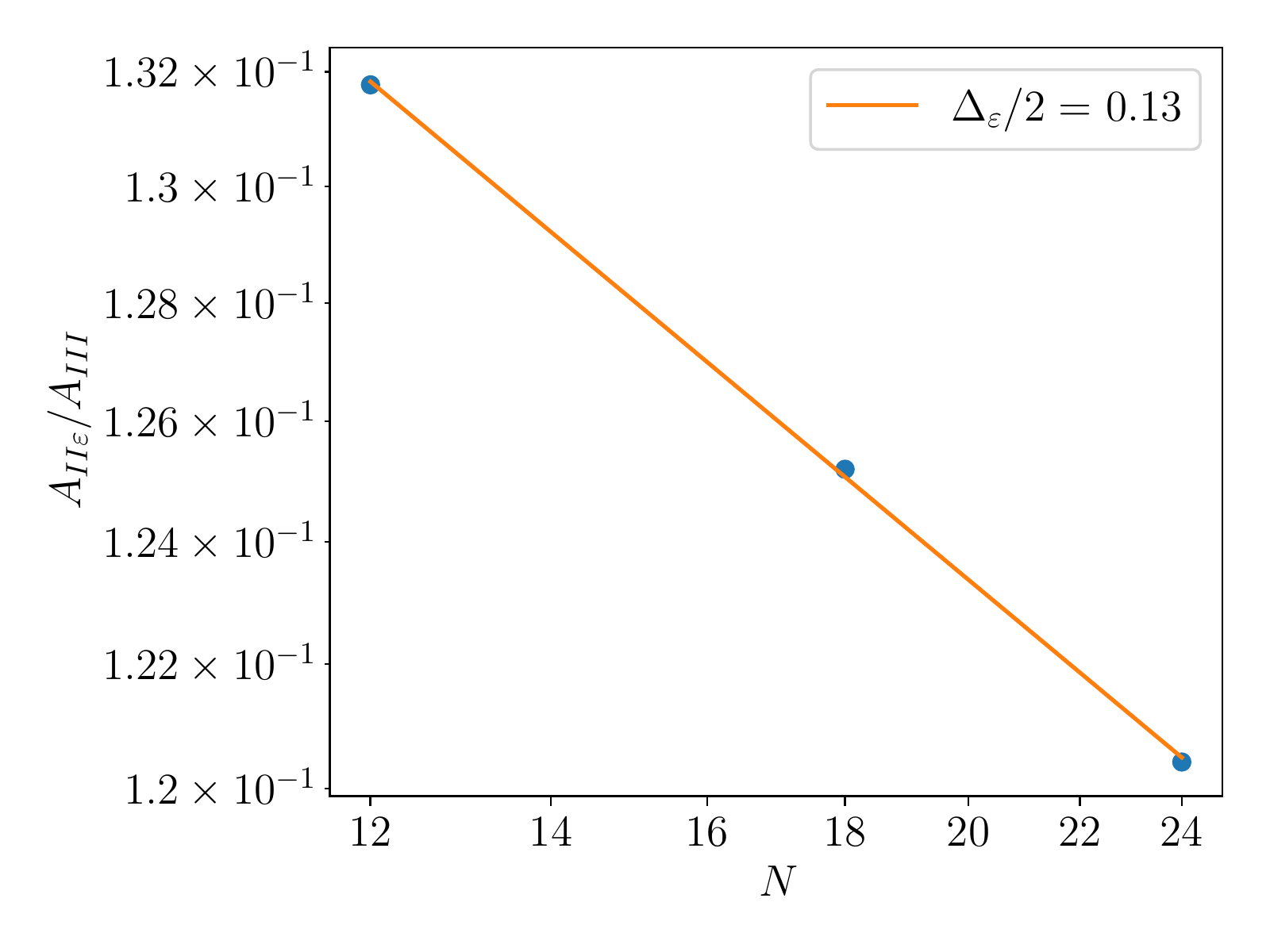}
    \caption{$A_{\bo\bo\varepsilon}/A_{\bo\bo\bo}$ for the Haagerup model. The states are taken from the Haagerup model with sizes $3n,3n$ and $N=6n$, where $n=2,3,4$.}
    \label{fig:IIep}
\end{figure}

Secondly, we compute the scaling dimension of the first spinless $\mathbb{Z}_3$ neutral excited state, which we denote as $|\varepsilon\rangle$. The relevant overlap is
\begin{eqnarray}
     \frac{A_{\bo\bo\varepsilon}}{A_{\bo\bo\bo}}\propto N^{-\Delta_{\varepsilon}/2}.
\end{eqnarray}
As shown in Fig.~\ref{fig:IIep}, we obtain $\Delta_{\varepsilon}\approx 0.26$. This agrees with the energy spectrum with reasonable accuracy.

\begin{figure}
    \centering
    \includegraphics[width=0.9\linewidth]{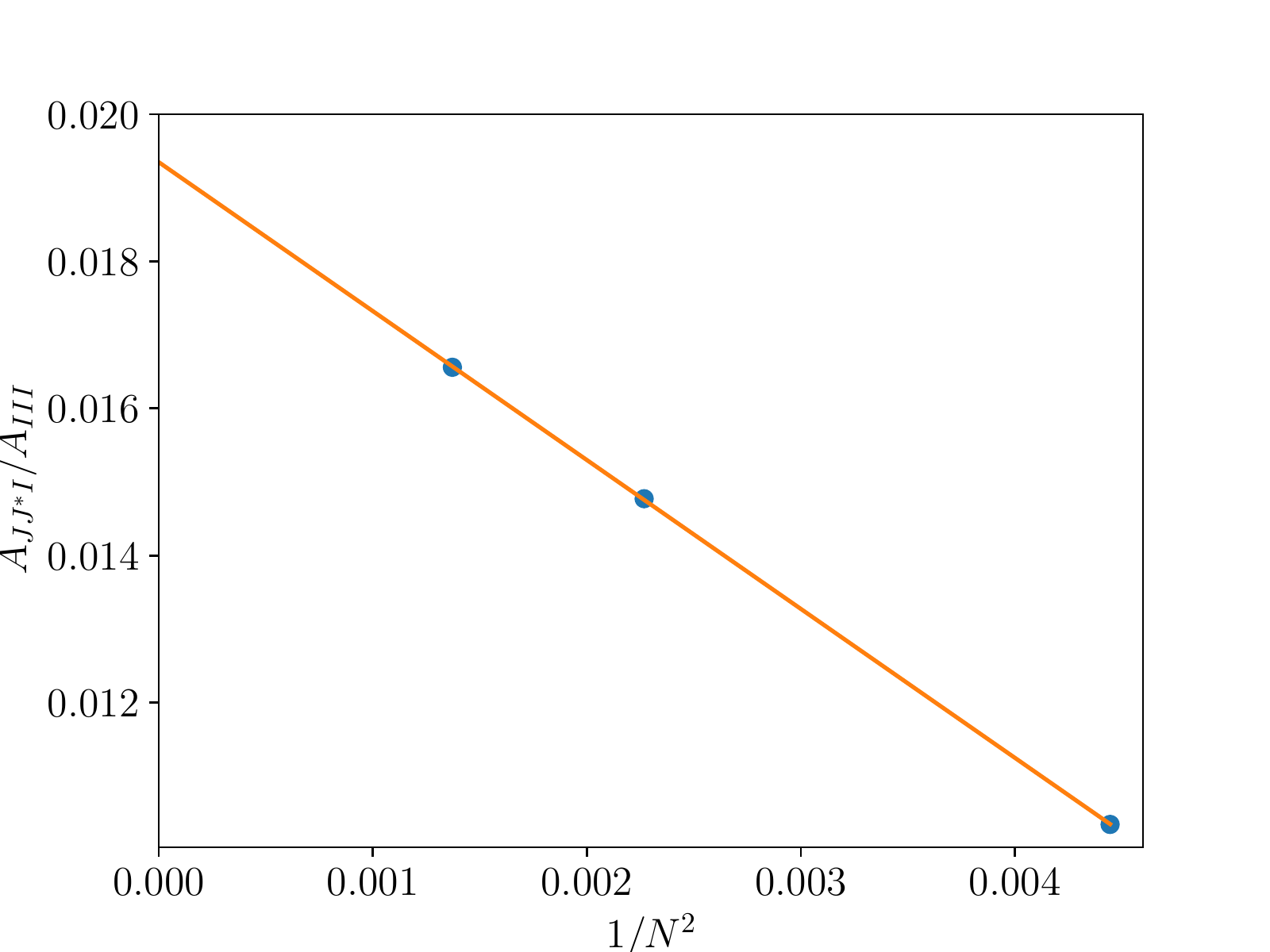}
    \caption{$A_{JJ^{*}\bo}/A_{\bo\bo\bo}$ for the Haagerup model. In $A_{JJ^{*}\bo}$ the three states are taken from the Haagerup model with sizes $3n+1,3n+2$ and $N=6n+3$. In $A_{\bo\bo\bo}$, the three states are taken from the Haagerup model with sizes $3n,3n+3$ and $N=6n+3$. The three points in the plot correspond to $n=2,3,4$.}
    \label{fig:JJI}
\end{figure}

Thirdly, we compute the $A_{JJ^*\bo}$ where $|J\rangle$ is the lowest eigenstate with $s_J=1$ at size $N=3n+1$. Similarly, $J^{*}$ is the lowest eigenstate with $s_{J^{*}}=1$ at size $N=3n+2$. Suppose the operator $J$ and $J^{*}$ are chiral currents with scaling dimension $1$, then they would be Kac-Moody descendants of $\bo$, and the wavefunction overlap $A_{JJ^*\bo}$ would only involve one fusion channel in the cyclic orbifold,
\begin{eqnarray}
     [(\bo,\bo)_s]\times [\tau_{(\bo,\hat{0})}]= [\tau_{(\bo,\hat{0})}]. 
\end{eqnarray}
This implies that the finite-size corrections to $A_{JJ^*\bo}$ would come from descendants of $\tau_{(\bo,\hat{0})}$, thus we would expect
\begin{equation}
\label{eq:JJI_conjecture}
     \frac{A_{J J^*\bo}}{A_{\bo\bo\bo}} = \frac{1}{16} +O(N^{-2})
\end{equation}
similar to the XXZ model considered in Sec.~\ref{sec:Ising_main}. However, our numerical result shown in Fig.~\ref{fig:JJI} violates Eq.~\eqref{eq:JJI_conjecture} as the extrapolation gives $A_{J J^*\bo}/{A_{\bo\bo\bo}}\approx 0.02$. 
Therefore we can conclude that $J$ is not a current. The scaling dimension can be estimated by
\begin{equation}
    \Delta_{J} = -\frac{1}{4}\log_{2}  \frac{A_{J J^*\bo}}{A_{\bo\bo\bo}},
\end{equation}
which gives $\Delta_J\approx 1.4$. Considering the error in finite size extrapolations, we conclude that $\Delta_{J}\leq 1.4$.


\section{Discussion}
\label{sec:discussion}

In this work we have studied wavefunction overlaps $A_{\alpha\beta\gamma} = \langle \phi_{\gamma}|\phi_{\alpha}\phi_{\beta}\rangle$ of a critical quantum spin chain, where the three states are primary states of the spin chain. Through a conformal mapping from the three-sided cylinder to the complex plane, we derive the relation between OPE coefficients $C_{\alpha\beta\gamma}$ and the ratio of wavefunction overlap $A_{\alpha\beta\gamma}/A_{\bo\bo\bo}$. 
In order to study the finite-size corrections, we rewrite $A_{\alpha\beta\gamma}$ as a path integral of the cyclic orbifold on the cylinder, with twist operator insertion at the branch points. We then show that finite-size correction exponents $p_{\alpha\beta\gamma}$ are universal, and are completely determined by the orbifold operator content. As a benchmark, we extracted OPE coefficients $C_{\alpha\beta\gamma}$ and finite-size correction exponents $p_{\alpha\beta\gamma}$ for the Ising model and the XXZ model and found that they agree with the prediction using the cyclic orbifold.

We then illustrate an application of our method to a less-known CFT. We have computed the wavefunction overlaps for the Haagerup model, and found that it has central charge $c\approx 2.1$, lowest spinless primary operator with dimension $\Delta_{\varepsilon}\approx 0.3$ and lowest spin-1 primary operator with dimension $\Delta_{J}\leq 1.4$. The fact that $A_{JJ^{*}\bo}/A_{\bo\bo\bo}$ is not close to $1/16$ provides negative evidence against the conjecture that $J$ is a current operator, so the Haagerup CFT is likely more complicated than the free boson $\mathbb{Z}_3$ orbifold.  

As noted in Ref.~\cite{2021arXiv210809366Z}, the overlap $A_{\alpha\beta\gamma}$ can be viewed as a wavefunction of a multiboundary wormhole state. In this work we interpret the same overlap as a three-point correlation function involving twist operators in the cyclic orbifold. It would be interesting for future work to explore the holographic interpretation of twist operators and cyclic orbifold. 

Given three primary states, it is possible to construct other overlaps that are inequivalent to the three-sided cylinder, but nevertheless also contain similar universal information of the CFT. One useful example is given in Ref.~\cite{liu2021multipartitioning}, where the authors consider the vertex state that describes a trisection of a two-dimensional chiral topological phase 
(see Appendix \ref{app:vertex} for details). 
It is shown in the appendix that conformal data, such as central charge and OPE coefficients, can be extracted from the wavefunction of the vertex state, with finite-size corrections that appear universal. It is an open question whether the finite-size corrections to these wavefunction overlaps are indeed universal, 
and if so, what operators are responsible. More generally, given a bunch of primary states, it is possible to construct many different overlaps, which may correspond to multi-point correlation functions of the CFT. The finite-size corrections remain elusive at this point. 

In conclusion, we have developed a method based on wavefunction overlaps that 
compute conformal data solely from low-energy eigenstates of a critical quantum spin chain. 
We have shown that the finite-size corrections are universal, which provides consistency conditions on the wavefunction overlaps at finite sizes. It is interesting both practically and conceptually. Practically, the method is useful for identifying new CFTs realized in lattice models without too much prior knowledge. 
Conceptually, the method relates wavefunction overlaps to operator fusion in the cyclic orbifold, which may be easily generalized.

\acknowledgements
We acknowledge 
useful discussions with Ramanjit Sohal,
Yuji Tachikawa, Masaki Tezuka, and Kantaro Ohmori.
YZ acknowledges useful discussions with Qi Hu and Guifre Vidal. YZ is supported by the Q-FARM fellowship at Stanford University. SR is supported by the National Science Foundation under 
Award No.\ DMR-2001181, and by a Simons Investigator Grant from
the Simons Foundation (Award No.~566116).

\bibliography{main_ref}

\newpage
\appendix

\section{Fusion rules of cyclic orbifold from modular transformations}
\label{app:modular_trans}

We first recall basic aspects of modular invariance which are crucial to the consistency of a CFT defined on a torus. A torus is characterized by two period vectors $w_1$ and $w_2$, which can be represented as complex numbers. Scale invariance of CFT implies that the partition function only depend on the modular parameter $\tau = w_2/w_1$. There are two basic operations $\mS:\tau\ra -1/\tau$ and $\mT:\tau\ra\tau+1$ that generate equivalent descriptions of the same torus. Therefore, the two operations generate the modular group under which the CFT partition function is invariant.

Modular invariance puts strong constraints on the operator content of the CFT. To see this, we first recall that the partition function is bilinear in the character of conformal towers,
\begin{equation}
    \chi_{a}(\tau) = \tr_a \, q^{L_0-c/24},
\end{equation}
where $q=\exp(2\pi i \tau)$, and $a$ is the label of primary operator. Expanding the character in the power of $q$ gives degeneracy of the conformal tower at each level of descendants. In a diagonal theory, the partition function is simply
\begin{equation}
    Z(\tau) = \sum_{a} |\chi_{a}(\tau)|^2,
\end{equation}
where the summation is finite for minimal model
\footnote{For rational CFT, the partition function can be organized into finite summation of extended characters. See the example of compactified boson 
(the XXZ model in the main text).}, 
for example, the Ising CFT. Modular invariance then demands that
\begin{eqnarray}
     \chi_{a}(-1/\tau) &=& \sum_{b}\mS_{ab} \chi_{b}(\tau), \\
     \chi_{b}(\tau+1) &=& \sum_{b}\mT_{ab} \chi_{b}(\tau), 
\end{eqnarray}
where $\mS$ and $\mT$ are unitary matrices that form a representation of the modular group. One can further show that 
\begin{equation}
    \mT_{ab} = e^{2\pi i (h_a-c/24)} \delta_{ab}
\end{equation}
is diagonal with $h_a=\Delta_a/2$ and
\begin{equation}
    \mS_{ab} = \mS_{ba}
\end{equation}
is symmetric. The celebrated Verlinde formula relates the fusion rule to the modular matrices,
\begin{equation}
    \mN_{i,j,k} = \sum_{m} \frac{\mS_{im} \mS_{jm} \mS^{*}_{mk}}{S_{\bo m}}.
\end{equation}
Next, we review how characters of the orbifold theory can be constructed from the original theory and use the Verlinde formula to derive the fusion rules.

The characters in the orbifold theory are related to the characters in the original theory via:
\begin{equation}
\begin{aligned}
&\chi_{(\alpha,\beta)}(\tau)=\chi_{\alpha}(\tau) \chi_\beta(\tau)\\
&\chi_{(\alpha,\alpha)_\phi}(\tau)=\frac{1}{2}\left[\chi_\alpha^2(\tau)+e^{i\pi\phi}\chi_\alpha(2\tau)\right]\\
&\chi_{(\alpha,\hat{\psi})}(\tau)=\frac{1}{2}\left[\chi_\alpha(\frac{\tau}{2})+e^{i\pi \psi}T_\alpha^{-1/2}\chi_\alpha(\frac{\tau}{2}+\frac{1}{2})\right].
\end{aligned}
\label{eqn:characters}
\end{equation}
Here $T_\alpha = \exp{(2\pi i (h_\alpha-c/24))}$ is the diagonal matrix element in $\mT$ matrix, $\phi$ can take $0$ (symmetric sector, denoted as $s$ in the main text) or $1$ (antisymmetric sector, denoted as $a$ in the main text), and $\psi$ can take $0$ or $1$. The partition function of the orbifold theory is:
\begin{equation}
    \begin{aligned}
         Z&=Z_{un}+Z_{tw},\\
         Z_{un}& = \sum_{\alpha<\beta}|\chi_{(\alpha,\beta)}(\tau)|^2+\sum_{\alpha,\phi}|\chi_{(\alpha,\alpha)_\phi}(\tau)|^2\\
         &=\frac{1}{2}\sum_{\alpha\beta}|\chi_\alpha(\tau)|^2|\chi_\beta(\tau)|^2+\frac{1}{2}\sum_\alpha|\chi_\alpha(2\tau)|^2,\\
         Z_{tw}&=\sum_{\alpha,\psi}|\chi_{(\alpha,\hat{\psi})}(\tau)|^2\\&= \frac{1}{2}\sum_\alpha(|\chi_\alpha(\frac{\tau}{2})|^2+|\chi_\alpha(\frac{\tau}{2}+\frac{1}{2})|^2).
    \end{aligned}
\end{equation}
We see that since the characters of $\phi_{(\alpha,\alpha)_{s/a}}$ include $\chi_\alpha(2\tau)$, it is necessary to include the twisted sector with characters $\chi_\alpha(\tau/2)$ and $\chi_\alpha(\tau/2+1/2)$ such that the partition function is modular invariant. 

For the purpose of computing wavefunction overlap, we need to use the following two orbifold fusion coefficients: $\mN_{(\alpha,\beta),{(\gamma,\hat{\psi})},{(\delta,\hat{\chi})}}$, $\alpha\neq \beta$ and $\mN_{(\alpha,\alpha)_s,{(\gamma,\hat{\psi})},{(\delta,\hat{\chi})}}$, where $\psi,\chi$ take value in $\lbrace 0,1\rbrace$. 

We now derive the Eq.\ (\ref{eqn:fusion_orbifold_1}), (\ref{eqn:fusion_orbifold_2}) in the main text. Using the Verlinde formula, the first fusion coefficient can be expressed as:
\begin{equation}
\begin{aligned}
    &\mN_{(\alpha,\beta),{(\gamma,\hat{\psi})},{(\delta,\hat{\chi})}}\\
    &=\sum_m \frac{\mS_{(\alpha,\beta),m} \mS_{(\gamma,\hat{\psi}),m}\mS^*_{m,(\delta,\hat{\chi})}}{\mS_{(\bo,\bo)_s,m}}\\
    &=\sum_{\eta, \phi}\frac{\mS_{(\alpha,\beta),(\eta,\eta)_\phi} \mS_{(\gamma,\hat{\psi}),(\eta,\eta)_\phi}\mS^*_{(\eta,\eta)_\phi,(\delta,\hat{\chi})}}{\mS_{(\bo,\bo)_s,(\eta,\eta)_\phi}}\\
    &=\sum_{\eta,\phi} \frac{\left(\mS_{\alpha\eta}\mS_{\beta\eta}\right)\left(\frac{1}{2}e^{i\pi\phi}\mS_{\gamma\eta}\right)\left(\frac{1}{2}e^{-i\pi\phi}\mS^*_{\eta\delta}\right)}{\frac{1}{2}\mS_{\bo\eta}^2}\\
    & = \sum_{\eta} \frac{\mS_{\alpha\eta}\mS_{\beta\eta}\mS_{\gamma\eta}\mS^*_{\eta\delta}}{\mS_{\bo\eta}^2}.
\end{aligned}
\end{equation}
From the second to the third line, 
we use the relation between 
the orbifold modular $\mS$ matrix and the original $\mS$ matrix, 
which can be derived using the orbifold character Eq.\ \eqref{eqn:characters} as in \cite{borisov1998systematic}\footnote{Note that there is a typo in the fusion coefficient $\mN$ derived in \cite{borisov1998systematic}, where the conjugate sign is missing.}.
For the case where $\mS$ is real 
(e.g., the Ising CFT), 
this still gives the correct result; while for the general case where $\mS$ is complex 
(e.g., the compactified boson rational CFT), this would change the final expression. Only when $\mS$ is real 
can we decompose $\mN_{(\alpha,\beta),{(\gamma,\hat{\psi})},{(\delta,\hat{\chi})}}$ 
into the product of $\mN$ 
in the original CFT. 
In short, the fusion rule of the orbifold CFT is fully determined by the modular matrices of the original theory.

Similarly, the second fusion coefficient can be written as:
\begin{widetext}
\begin{equation}
\begin{aligned}
&\mN_{(\alpha,\alpha)_s,{(\gamma,\hat{\psi})},{(\delta,\hat{\chi})}}\\
&=\sum_m \frac{\mS_{(\alpha,\alpha)_s,m} \mS_{(\gamma,\hat{\psi}),m}\mS^*_{m,(\delta,\hat{\chi})}}{\mS_{(\bo,\bo)_s,m}}\\ 
& = \sum_{\eta,\phi}\frac{\mS_{(\alpha,\alpha)_s,(\eta,\eta)_\phi} \mS_{(\gamma,\hat{\psi}),(\eta,\eta)_\phi}\mS^*_{(\eta,\eta)_\phi,(\delta,\hat{\chi})}}{\mS_{(\bo,\bo)_s,(\eta,\eta)_\phi}} + \sum_{\eta,\xi}\frac{\mS_{(\alpha,\alpha)_s,(\eta,\hat{\xi})} \mS_{(\gamma,\hat{\psi}),(\eta,\hat{\xi})}\mS^*_{(\eta,\hat{\xi}),(\delta,\hat{\chi})}}{\mS_{(\bo,\bo)_s,(\eta,\hat{\xi})}}\\
&=\sum_{\eta,\phi} \frac{\left(\frac{1}{2}\mS^2_{\alpha\eta}\right)\left(\frac{1}{2}e^{i\pi\phi}\mS_{\gamma\eta}\right)\left(\frac{1}{2}e^{-i\pi\phi}\mS^*_{\eta\delta}\right)}{\frac{1}{2}\mS_{\bo\eta}^2} + \sum_{\beta,\xi} \frac{\left(\frac{1}{2}\mS_{\alpha\eta} \right)\left(\frac{1}{2}e^{i\pi(\psi+\xi)}\mP_{\gamma\eta}\right)\left(\frac{1}{2}e^{-i\pi(\xi+\chi)}\mP^*_{\eta,\delta}\right)}{\frac{1}{2}\mS_{\bo\eta}}\\
&=\frac{1}{2}\sum_\eta \frac{\mS_{\alpha\eta}^2 \mS_{\gamma\eta}\mS^*_{\eta\delta}}{\mS^2_{\bo\eta}}+\frac{1}{2}e^{i\pi(\psi+\chi)}\sum_\eta\frac{\mS_{\alpha\eta}\mP_{\gamma\eta}\mP^*_{\eta,\delta}}{\mS_{\bo,\eta}},
\end{aligned}
\end{equation}
\end{widetext}
where $\mP=\mT^{1/2}\mS \mT^2\mS \mT^{1/2}$.

\section{Wavefunction overlaps for antiferromagnetic Ising model}
\label{app:antiferro}


In the main text, we consider the ferromagnet Ising model, where $\bo,\sigma,\epsilon$ are the primary operators when taking periodic boundary condition (PBC), and $\mu,\psi,\bar{\psi}$ appears in anti-periodic boundary condition (APBC). There is no distinction of low-energy spectrum for even and odd number of sites, and all the primary operators are inside the light cone at $P=0$.

On the other hand, the low-energy spectrum of the antiferromagnetic Ising model depends on even or odd number of sites. 
Furthermore, there exists lightcones centered at $P=0$ as well as $P=\pi$. The Haagerup model is more similar to the antiferromagnetic Ising model, in the sense that (1) the operator content changes with different system sizes $N=3n,3n+1,3n+2$; and (2) there appear three light cones which are located at $P=0,P=\pi/3,P=-\pi/3$ if $N=3n$. For this reason, we study the wavefunction overlap for the antiferromagnetic Ising model in this appendix, with an emphasis on the case where low-energy eigenstates with odd number of sites fuse into low-energy eigenstates with even number of sites. This provides an analogy to what we did for the Haagerup CFT.

The operator content of the antiferromagnetic Ising model is summarized below, 
for both boundary conditions (PBC/APBC) and for both types of lattice site (even/odd):
\begin{equation}
    \begin{tabular}{c|c|c|c}
\hline
 & spin PBC & spin APBC & fermion\\
\hline
even &  $\mathbb{1}, \epsilon: P=0$ &  $\psi,\bar{\psi}: P =0$ & NS\\
&  $\sigma: P=\pi$ &  $\mu: P=\pi$ & R\\
\hline
odd &   $\psi,\bar{\psi}: P=\pi$ & $\mathbb{1}, \epsilon: P=\pi$ & NS\\
& $\mu: P=0$ &  $\sigma: P=0$ & R\\
\hline
\end{tabular}
\end{equation}

\begin{figure}[t]
\begin{subfigure}[b]{0.45\linewidth}
\centering
\includegraphics[width=\textwidth]{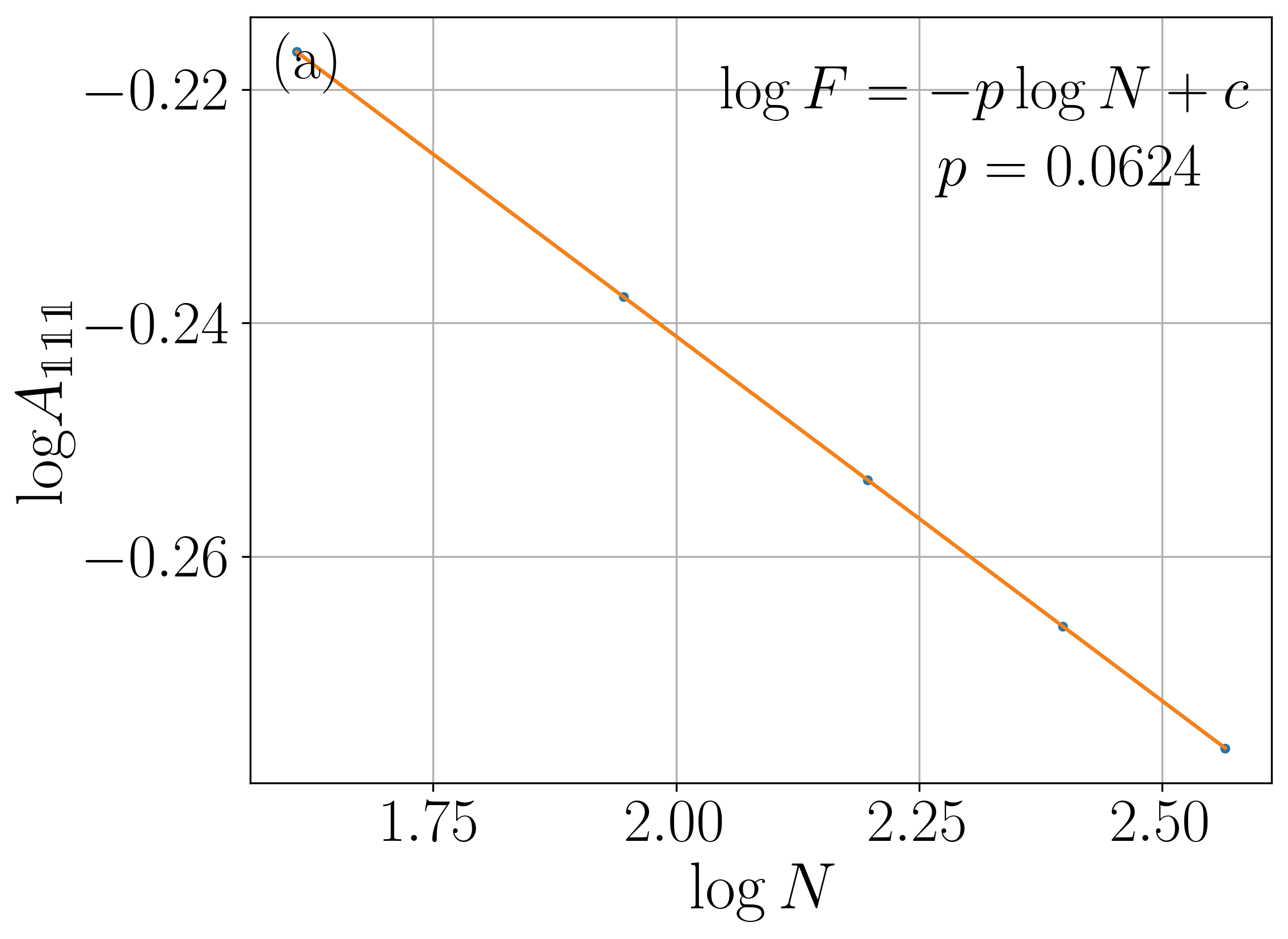}
\end{subfigure}
\begin{subfigure}[b]{0.45\linewidth}
\centering
\includegraphics[width=\textwidth]{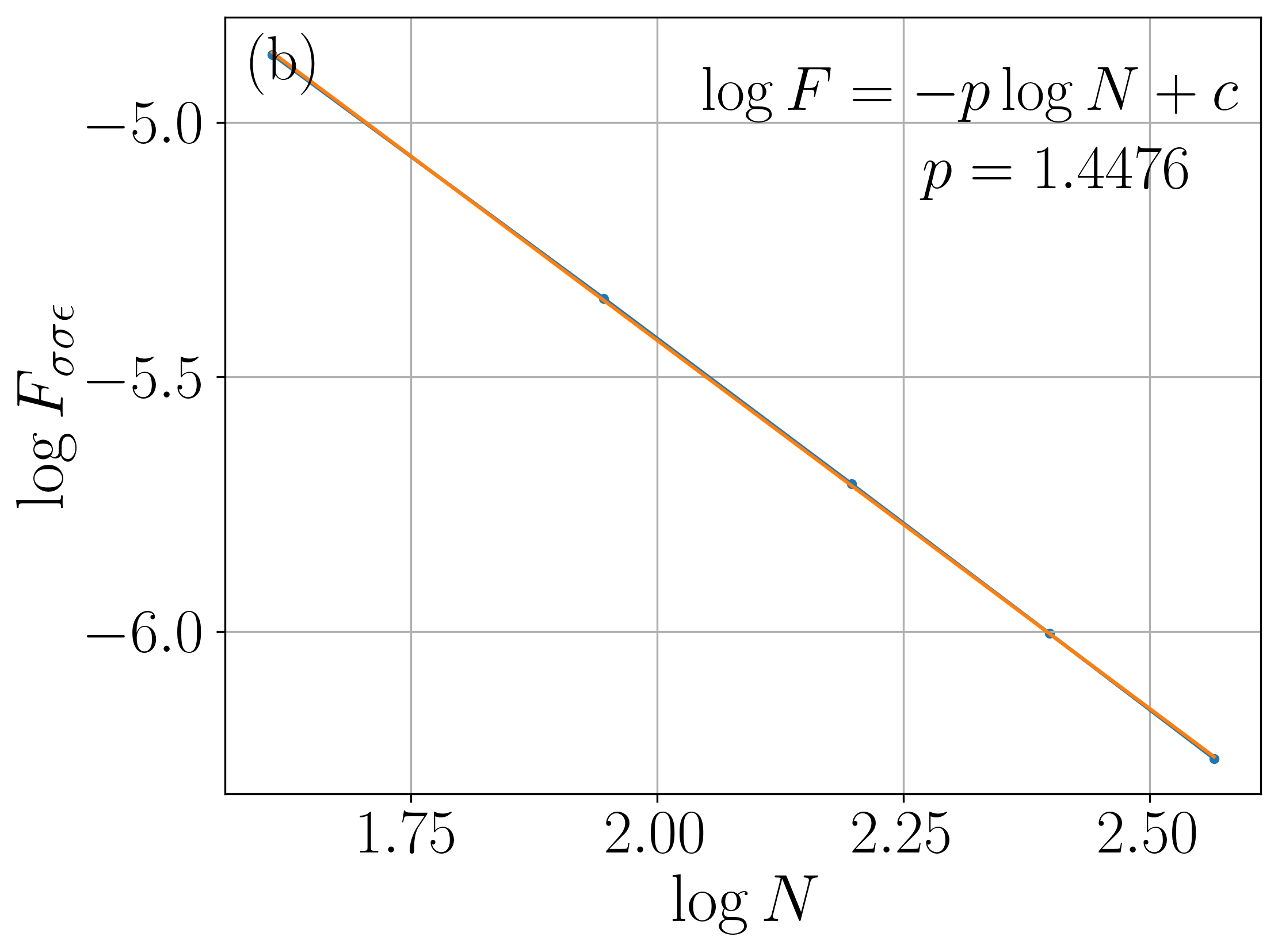}
\end{subfigure}
\begin{subfigure}[b]{0.45\linewidth}
\centering
\includegraphics[width=\textwidth]{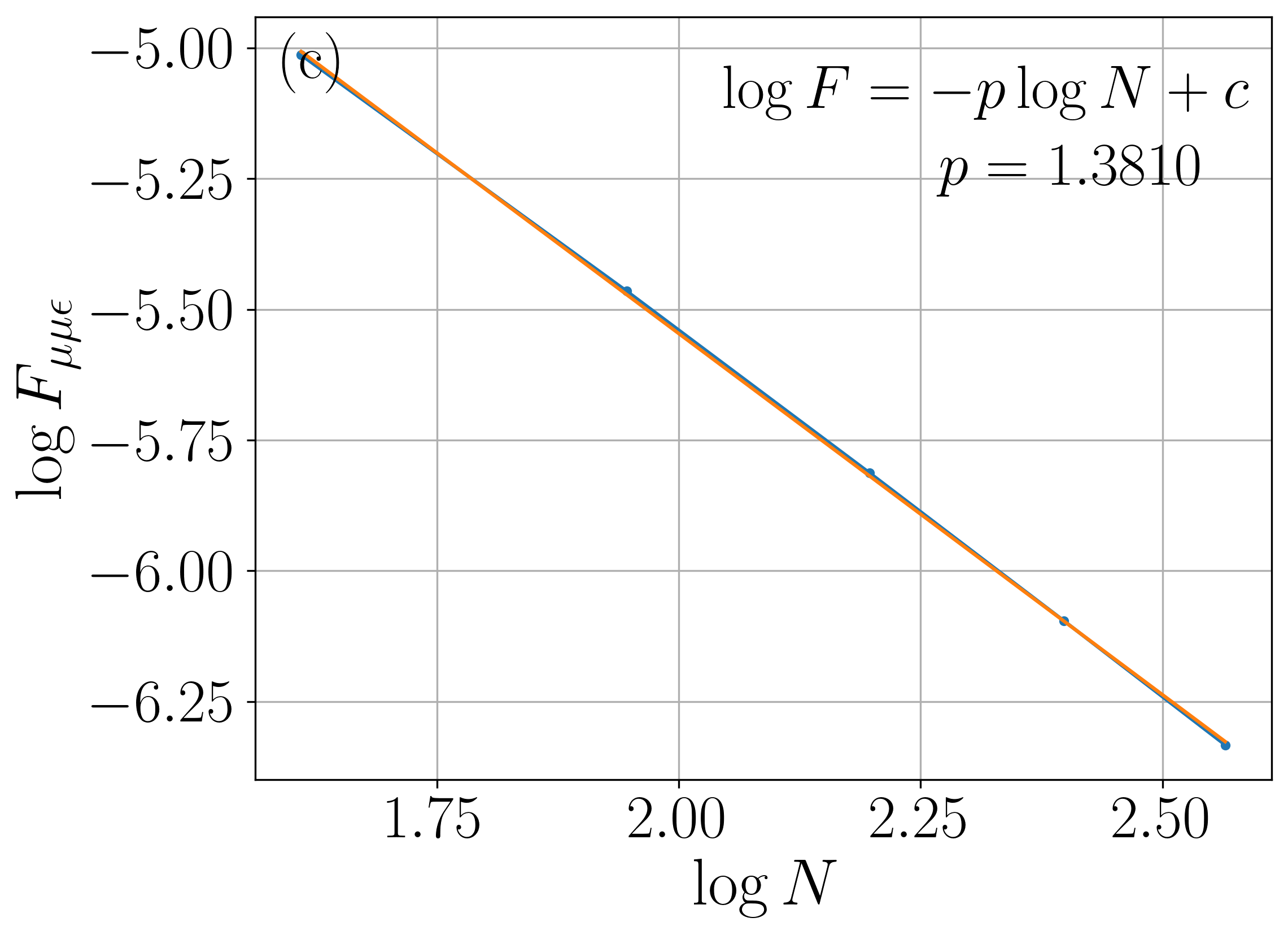}
\end{subfigure}
\begin{subfigure}[b]{0.45\linewidth}
\centering
\includegraphics[width=\textwidth]{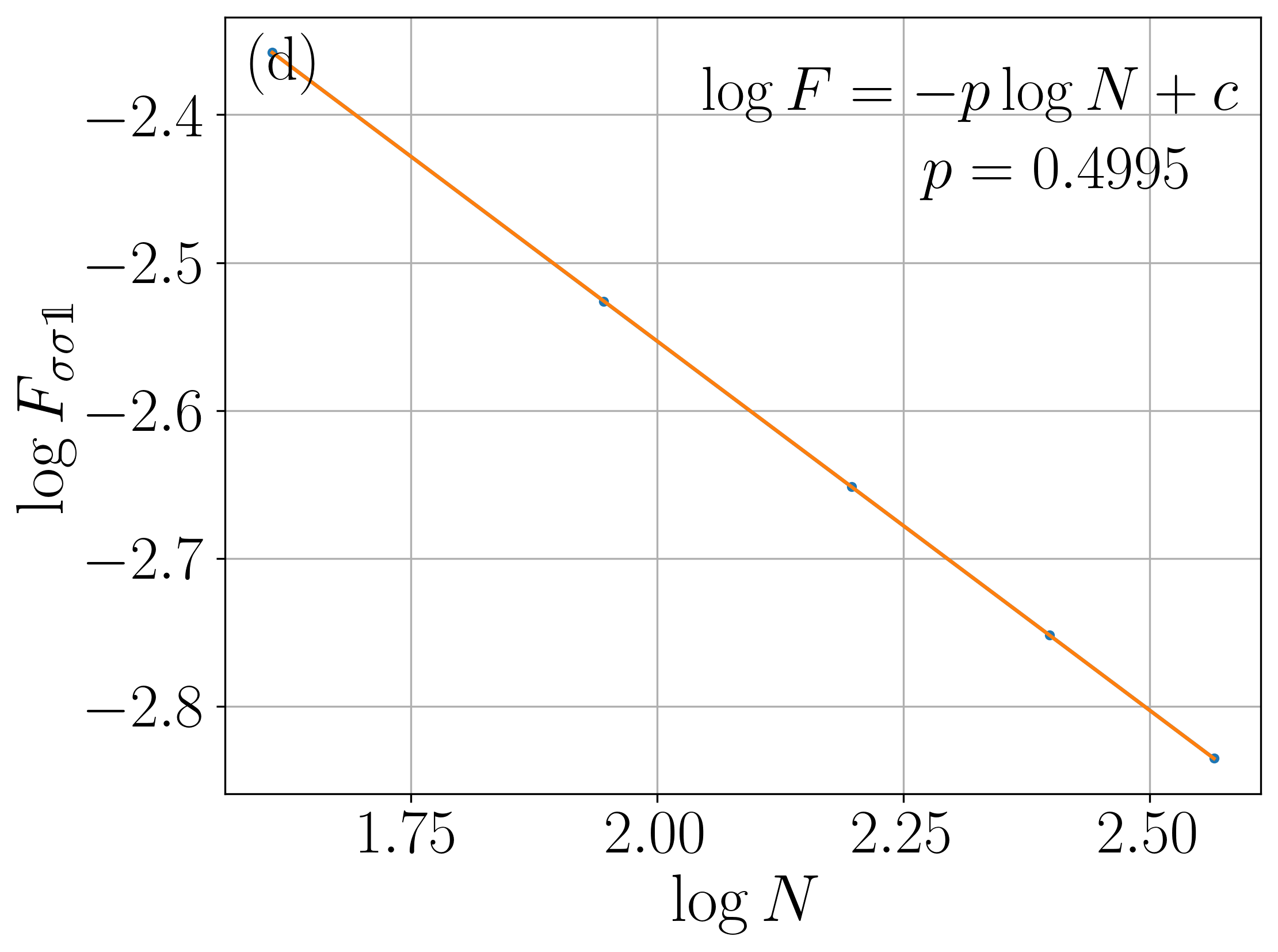}
\end{subfigure}
\begin{subfigure}[b]{0.45\linewidth}
\centering
\includegraphics[width=\textwidth]{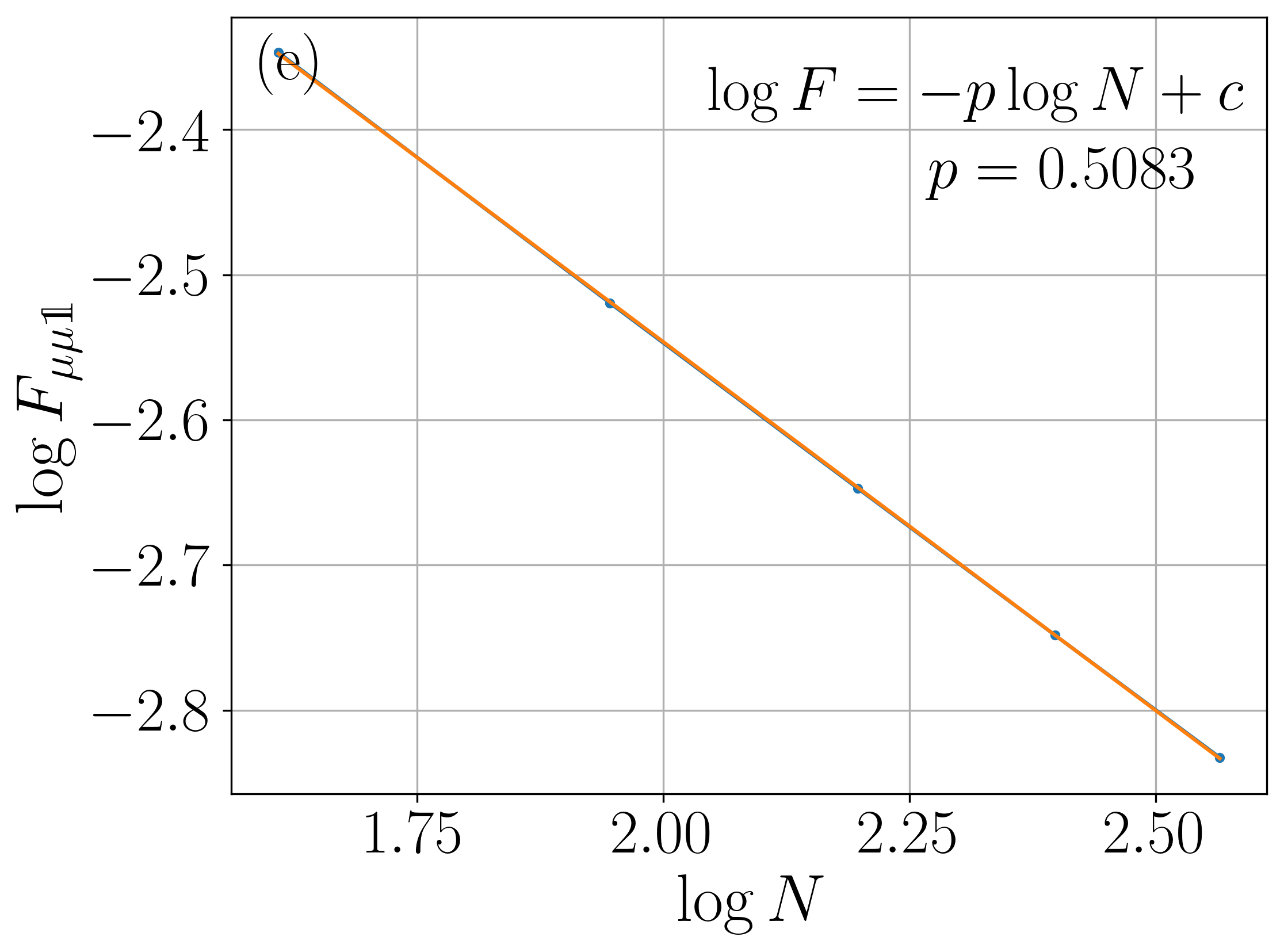}
\end{subfigure}
\begin{subfigure}[b]{0.45\linewidth}
\centering
\includegraphics[width=\textwidth]{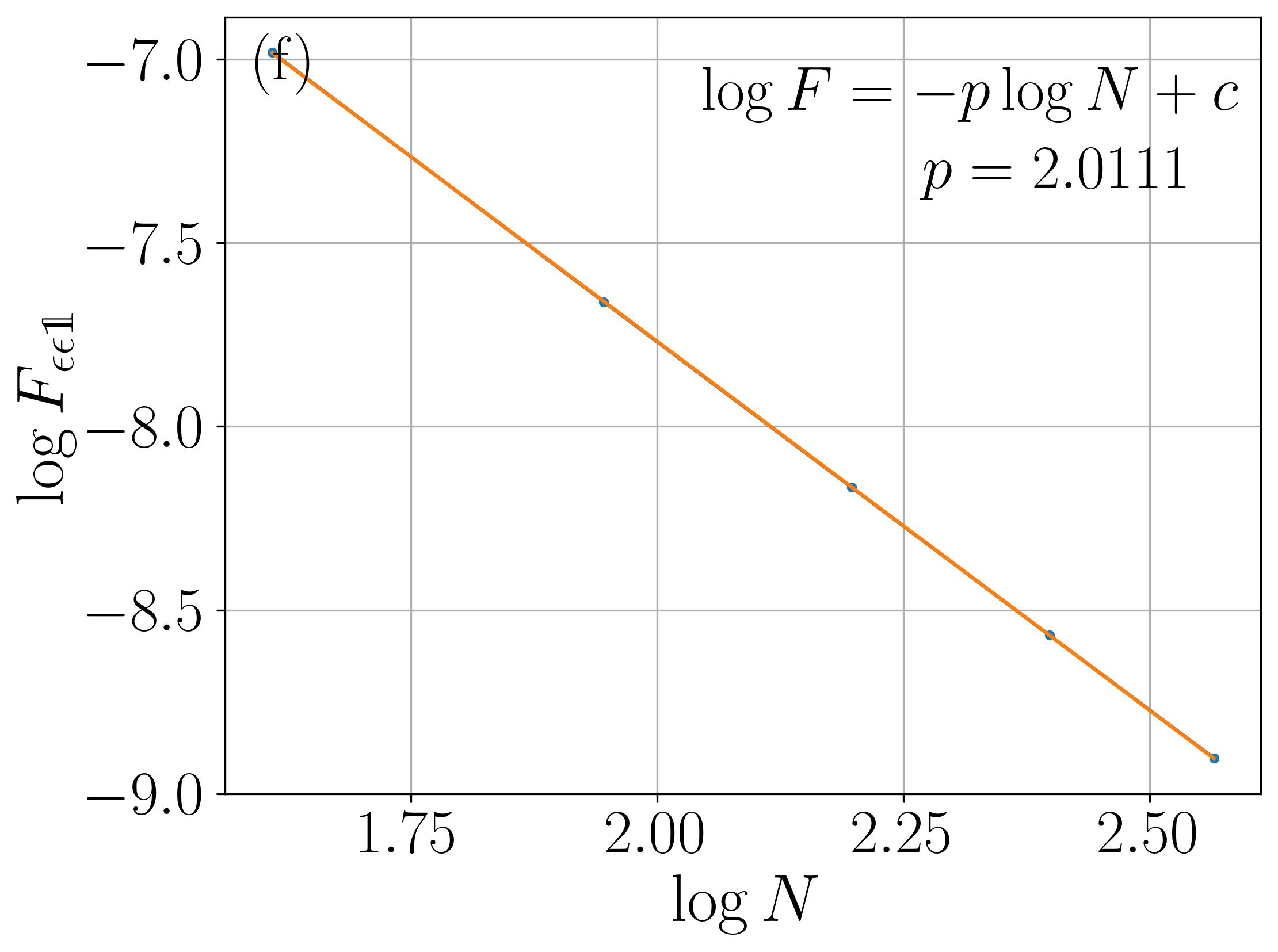}
\end{subfigure}
\caption{(a) Scaling of $A_{\bo\bo\bo}$, computed for $N_1=2N$, $N_2=2N$, and $N_3=4N$. (b-f) Scaling of $F_{\alpha\beta\gamma}$, where $A_{\alpha\beta\gamma}$ is computed for $N_1=2N+1$,$N_2=2N-1$, and $N_3=4N$. To simulate small system, $N$ takes value from 5 to 12. }
\label{fig:AFMIsing}
\end{figure}

A possible fusion may be, for example, obtaining $A_{\sigma\sigma\bo}$ by fusing $\sigma$ at size $N_1=2N+1$ and $\sigma$ at size $N_2=2N-1$ into $\bo$ at size $N_3=4N$. 
This is analog to what we did for the current operator in the Haagerup model. We then take the ratio $A_{\sigma\sigma\bo}/A_{\bo\bo\bo}$ where the $\bo$ in the denominator is taken from $N_1=N_2=2N$ and $N_3=4N$.  For a better comparison with Haagerup model, we use the small systems, where $N$ takes value from 5 to 12, and show the results in Fig.\ \ref{fig:AFMIsing}. 

In Fig.\ \ref{fig:AFMIsing}(a), we show the scaling of $A_{\bo\bo\bo}$ computed using $N_1=2N,N_2=2N,N_3=4N$, where the theoretical derivation gives $A_{\bo\bo\bo}=a_{(\bo,\hat{0})} N^{-1/16}$. The fit gives the power $-0.0623$, which is within $1\%$ error even for such a small system. In Fig.\ \ref{fig:AFMIsing}(b-f), we show the finite-size correction $F_{\alpha\beta\gamma}$ for the five different channels $\sigma\sigma\epsilon,\mu\mu\epsilon,\sigma\sigma \bo,\mu\mu \bo,\epsilon\epsilon \bo$, where $A_{\alpha\beta\gamma}$ is computed using $N_1=2N-1,N_2=2N+1,N_3=4N$, and $A_{\alpha\beta\gamma}^{(0)}$ is computed using Eq.~\eqref{eqn:A_full_expression} with different system size.  From the value of $\log F$, we see in (f) the numerical value approaches the theoretical value much quicker that (b,c); and (b,c) are quicker than (d,e). This is expected because $p=2$ for $\epsilon\epsilon\bo$, $p=\frac{3}{2}$ for $\sigma\sigma \bo$ and $\mu\mu\bo$; and $p=\frac{1}{2}$ for $\sigma\sigma\epsilon,\mu\mu\epsilon$. At this system size, the largest error for $p$ is $8\%$, and the error for $p$ gets smaller with increasing $N$. 


The above observation shows that the overlap result $A_{\alpha\beta\gamma}/A_{\bo\bo\bo}$ is indeed more accurate for larger $p_{\alpha\beta\gamma}$. 

The extraction for $p$ matches with the value in Table \ref{tab:Ising-finite} as we further increase the system size $N$. 

\section{Periodic uniform matrix product states}
\label{app:pumps}
For a quantum spin chains with $N$ spins and each spin with Hilbert space dimension $d$, a periodic uniform matrix product state (puMPS) is
\begin{equation}
    |\psi(A)\rangle = \sum_{\vec{s}}\tr(A^{s_1}A^{s_2}\cdots A^{s_N}) |\vec{s}\rangle,
\end{equation}
where $\vec{s}\equiv s_1 s_2\cdots s_N$, $s_i = 1,2,\cdots d$ and $A^{s}$ is a $D\times D$ matrix, with $D$ being the bond dimension. For a critical system, a puMPS can represent the ground state faithfully given that $D$ grows polynomially with $N$. The tensor $A$ can be found by optimizing the energy. 
An ansatz for excited states with momentum $p$ is given by the puMPS Bloch state
\begin{equation}
    |\phi_{p}(B;A)\rangle = \sum_{j} e^{ipj}\mathcal{T}^j \sum_{\vec{s}}\tr(B^{s_1}A^{s_2}\cdots A^{s_N})|\vec{s}\rangle,
\end{equation}
where $\mathcal{T}$ is the translation operator. It has been shown that this ansatz can capture all low-energy excitations given large enough bond dimensions. The tensor $B$ can be found by solving a generalized eigenvalue problem in the tangent space. The total cost for the ground state optimization and finding low-energy excitations is $O(ND^5)$ and $O(ND^6)$, respectively. Moreover, it is not hard to show that an anyon chain model of the form
\begin{equation}
    H = \sum_{j=1}^N \sum_{a,b=1}^d O^{a}_{j-1} h^{a,b}_{j} O^{b}_{j+1}
\end{equation}
can be represented as a matrix product operator (MPO) with bond dimension $D_{\mathrm{MPO}}=2d+2$, where the MPO tensor is 
\begin{equation}
M_{\mathrm{MPO}}=
    \begin{bmatrix}
    \bo & & & \\
    O^T & & & \\
      &h & & \\
      &  & O & \bo
    \end{bmatrix}
\end{equation}
where $O = (O_1, O_2,\cdots, O_d)$ is a $1\times d$ block and $h$ is a $d\times d$ block with entries $h^{a,b}$. 
For the Haagerup model $d=6$ and $O^{a}$ is a projector onto a specific anyon subspace, and $h^{a,c}$ has matrix elements $h^{a,b}_{\alpha,\beta} = -F^{a\rho\rho}_{b \alpha \rho}F^{a\rho\rho*}_{b \beta \rho}$, where $\alpha,\beta=1,2,\cdots,d$ label physical indices. This MPO is useful for the algorithm of finding the excitations. 

Finally, given three puMPS Bloch states $|\Phi_{p_i}(B_i;A_i)\rangle$ with system sizes $N_i$ ($i=1,2,3$) and $N_1+N_2=N_3$, the wavefunction overlap $\langle\Phi_{p_3}(B_3;A_3)|\Phi_{p_1}(B_1;A_1)\Phi_{p_2}(B_2;A_2)\rangle$ can be computed with cost $O(D^2_3(N_1D^2_1+N_2D^2_2))$. 
\section{Finite-size corrections for general overlaps}
\label{app:generalized}
In the main text, we consider the finite-size correction to Eq.~\eqref{eq:ov3state_gen}, where $L_1=L_2$. The finite-size correction in this case is solvable using the cyclic orbifold. In the more general case where $L_1\neq L_2$, the path integral on Riemann surface $\Sigma$ is not the simple $\mathsf{N}=2$ cyclic orbifold. 

We numerically compute finite-size correction to wavefunction overlap in Table.\ \ref{tbl: diff_ratio}, for the three examples $L_2=2L_1$, $L_2=3L_1$ and $L_2=\frac{3}{2}L_1$. In all the cases, $A_{\bo\bo\bo}$ scales as $N^{-\frac{1}{16}}$, and the values of $p$ are close to either $\frac{1}{2}$ or $2$. It is worth noticing that for a given channel, the values of $p$ are the same for all the three choices in the $N\ra\infty$ limit. Note that all powers are still $p \in \{\Delta_{(\alpha,\hat{\chi})}-\Delta_{(\bo,\hat{0})}\}$, indicating that they are determined by the scaling dimension of the twist operator insertion at the singular point. However, not all $p$'s are the same as the $L_1=L_2$ case. For example, $p_{\sigma\sigma\epsilon}=1/2$ for $L_1\neq L_2$, as opposed to $3/2$ for $L_1= L_2$. This indicates that the wavefunction overlaps for $L_1\neq L_2$ are no longer three-point correlation functions of the cyclic orbifold, but more complicated correlation functions of the cyclic orbifold. 

(The case $L_2 = \frac{4}{3}L_1$ is also computed with the same trend. Not listed in the table.)

\begin{table}
    \centering
    \begin{tabular}{c|c|c|c|c|c|c}
    \toprule
     &  \multicolumn{6}{c}{$p$}\\
    \hline
       overlap &  \multicolumn{2}{c}{$L_2=2L_1$} & \multicolumn{2}{c}{$L_2=3L_1$} & \multicolumn{2}{c}{$L_2=\frac{3}{2}L_1$}\\
         \hline
    $\bo \times\bo\ra\epsilon$     & 0.4999& $\hspace{0.2em}\frac{1}{2}\hspace{0.2em}$ & 0.4996 & $\hspace{0.2em}\frac{1}{2}\hspace{0.2em}$ & 0.4999 & $\hspace{0.2em}\frac{1}{2}\hspace{0.2em}$  
    \\
    $\bo\times \epsilon\ra\bo$ & 0.4983 & $\frac{1}{2}$ & 0.4978 & $\frac{1}{2}$ & 0.4969 & $\frac{1}{2}$
    \\
    $\epsilon\times \epsilon\ra\bo$ & 2.0003 & 2 & 2.0010 & 2 & 2.0002 & 2
    \\
    $\epsilon\times \epsilon\ra\epsilon$ & 0.4991 & $\frac{1}{2}$ & 0.4974 & $\frac{1}{2}$ & 0.4994 & $\frac{1}{2}$
    \\
    $\bo\times \epsilon\ra\epsilon$ & 1.9946 & 2 & 1.9957 & 2 & 1.9828 & 2
    \\
    $\bo\times \sigma\ra\sigma$ & 0.4993 & $\frac{1}{2}$ & 0.4989 & $\frac{1}{2}$ & 0.4990 & $\frac{1}{2}$
    \\
    $\sigma\times \epsilon\ra\sigma$ & 0.4994 & $\frac{1}{2}$ & 0.4996 & $\frac{1}{2}$ & 0.4984 & $\frac{1}{2}$
    \\
    $\sigma\times \sigma\ra \bo$ & 0.5000 & $\frac{1}{2}$ & 0.4999 & $\frac{1}{2}$ & 0.5000 & $\frac{1}{2}$
    \\
    $\sigma\times \sigma\ra\epsilon$ & 0.4827 & $\frac{1}{2}$ & 0.4909 & $\frac{1}{2}$ & 0.4057 & $\frac{1}{2}$\\
   \hline\hline
    \end{tabular}
    \caption{Finite-size correction to the wavefunction overlap for three more general cases $L_2=2L_1,L_2=3L_1$, and $L_2=\frac{3}{2}L_1$. The value of system size $N_1$ is taken in the range $[200,500]$. $A_{\bo\bo\bo}$ scales as $N^{-1/16}$. The values for $p$ are close to $1/2$ and $2$. For $\sigma\times \sigma\ra\epsilon$ channel in the case  $L_2=\frac{3}{2}L_1$, the value will go to 0.5 if we further increase the system size. }
    \label{tbl: diff_ratio}
\end{table}

\section{OPE from vertex state}
\label{app:vertex}
\begin{figure}[tbh]
    \centering
    \includegraphics[width=\columnwidth]{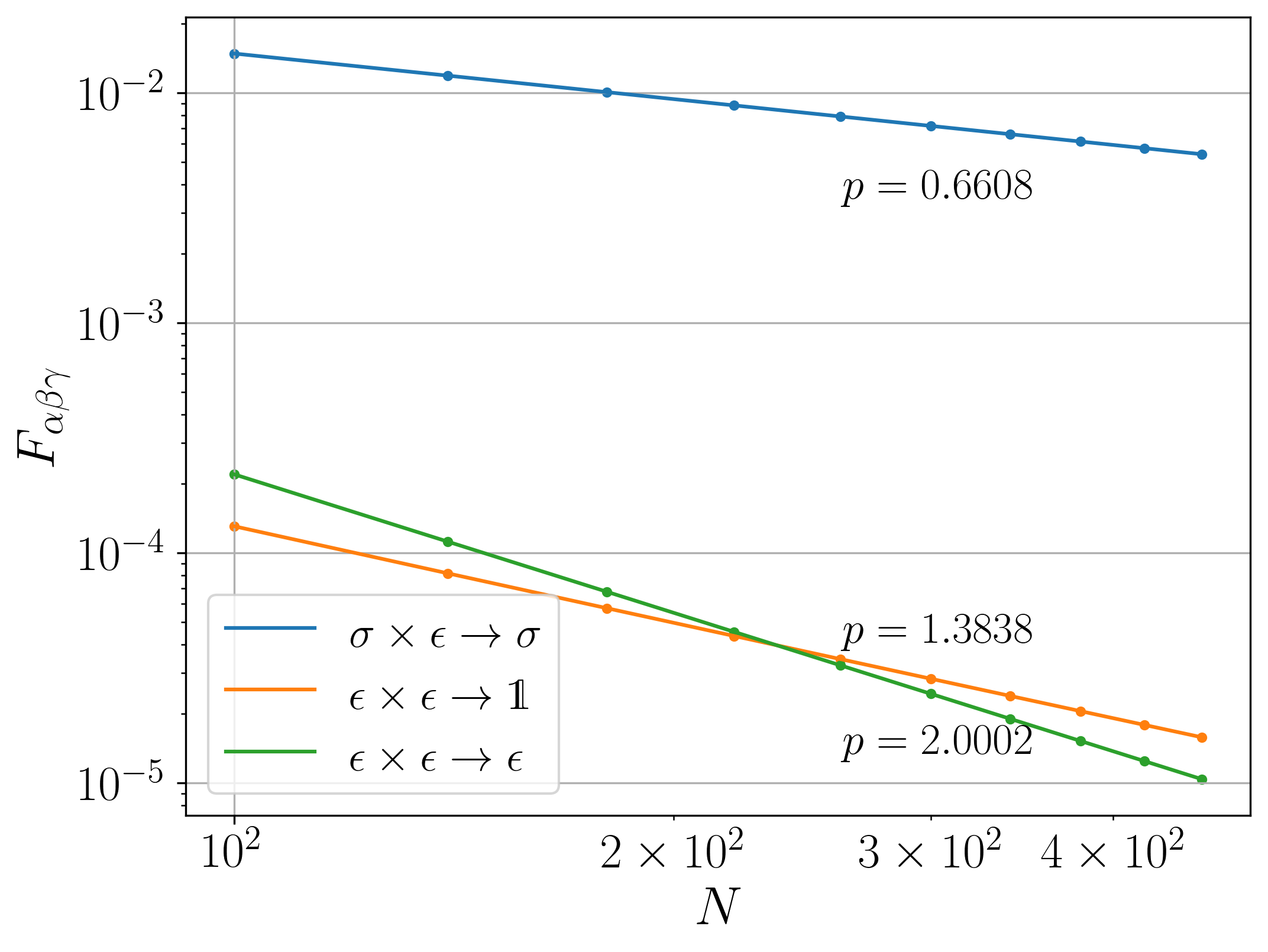}
    \caption{Finite-size correction of wavefunction overlap of Ising model using the vertex state , for system size $N\in[100,500)$.}
    \label{fig:Ising_twist}
\end{figure}

The extraction of OPE coefficient from wavefunction overlap can be applied to even more general cases. In Sec.\ \ref{sec:wavefunction_overlap}, we discuss the overlap configuration of three cylinders with circumference $L_1, L_2, L_3$ and $L_1+L_2=L_3$, where two cylinders ($L_1,L_2$) are joined into one cylinder ($L_3$). In this appendix, we consider a different configuration, where all three cylinders have the same circumference $L$, as shown in Fig.\ \ref{fig:Ising_twist}. This configuration was considered by some of the authors \cite{liu2021multipartitioning} in the context of vertex state in CFT.

\begin{figure}
    \centering
    \includegraphics[width=0.9\linewidth]{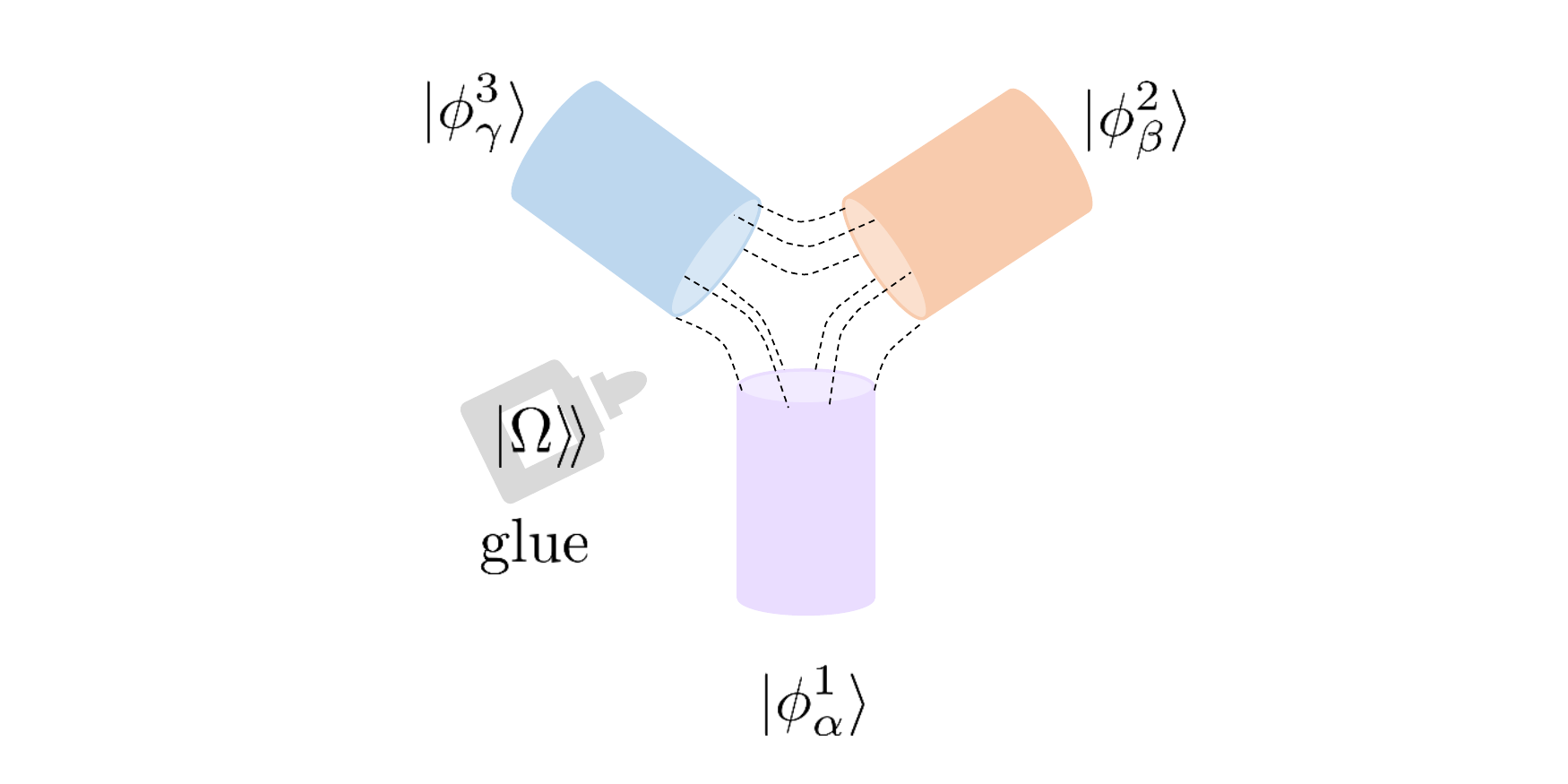}
    \caption{Wavefunction overlap of three cylinders with the same circumference. The maximally-entangled state $|\Omega\rangle\!\rangle$ acts as a ``glue" to form the path integral on the Riemann surface. }
    \label{fig:cartoon_twist}
\end{figure}

The wavefunction overlap in this configuration can be understood as follows. Let us define the maximally entangled state $|\Omega\rangle\!\rangle$ which can be regarded as a ``glue" (as illustrated in Fig.\ \ref{fig:cartoon_twist}). 
This is nothing but the vertex 
state studied in 
\cite{liu2021multipartitioning}.
Using $|\Omega\rangle\!\rangle$, we can glue the three half infinite cylinders into a junction shape of surface $\Sigma$. Denote the wavefunction overlap as:
\begin{eqnarray}
     A_{\alpha\beta\gamma}= \langle\!\langle \Omega|(|\phi^1_\alpha\rangle|\phi^2_\beta\rangle |\phi^3_\gamma\rangle)
\end{eqnarray}
Note that this is not the same overlap as appeared in the main text. After the conformal transformation $\omega(z)$, the path integral on this surface $\Sigma$ is related to the three-point function on the plane:
\begin{equation}
\begin{aligned}
   \frac{A_{\alpha\beta\gamma}}{A_{\bo\bo\bo}}
   &=\langle \phi^{1}_\alpha(-\infty_\alpha)\phi^{2}_\beta(-\infty_\beta)\phi^{3}_\gamma(-\infty_\gamma)\rangle_{\Sigma}\\
    & = \prod_i\left(\frac{d\omega_i}{dz}\right)^{\Delta_i}\langle \phi^1_\alpha(e^{i\pi/3})\phi^2_\beta(e^{-i\pi/3})\phi^3_\gamma(e^{-i\pi})\rangle_{\mathbb{C}},
\end{aligned}
\end{equation}
where the conformal transformation $\omega(z)$ maps $\Sigma$ to a complex plane:
\begin{equation}
    \omega_i=\omega_{i,0}\left(\frac{1+z}{1-z}\right)^{2/3},
\end{equation}
with $\omega_{1,0}=e^{i\pi/3},\omega_{2,0}=e^{-i\pi/3},\omega_{3,0}=e^{-i\pi}$, $z=e^{\tau+i\sigma}$. Plug in $d\omega/dz$ and the three-point correlation function on a plane, we obtain:
\begin{equation}
\begin{aligned}
     \frac{A_{\alpha\beta\gamma}}{A_{\bo\bo\bo}}
     &=\left(\frac{4}{3\sqrt{3}}\right)^{\Delta_\alpha+\Delta_\beta+\Delta_\gamma}C_{\alpha\beta\gamma}.
\end{aligned}
\label{eqn:vertex_ope}
\end{equation}

The above relation is checked numerically using ferromagnetic Ising model with accurate agreement for extracted OPE. 
In this configuration, however, the form of finite-size correction to Eq.~\eqref{eqn:vertex_ope} is different, because the cyclic orbifold picture does not apply to this case. We show the numerical obtained $p$ in Table \ref{tab:Ising_vertex} using $N\in [100,500)$ and the plot for three channels in Fig.~\ref{fig:Ising_twist}, where:
\begin{equation}
\begin{aligned}
    F_{\alpha\beta\gamma}&=\frac{A_{\alpha\beta\gamma}}{A_{\bo\bo\bo}}-\left(\frac{4}{3\sqrt{3}}\right)^{\Delta_\alpha+\Delta_\beta+\Delta_\gamma}C_{\alpha\beta\gamma}\\
    &=\text{const}\cdot N^{-p_{\alpha\beta\gamma}}.
\end{aligned}
\end{equation}
 The values of $p$ are close to three values: $0.66, 1.38, 2.00$. This indicates that finite-size corrections for the overlap with the vertex state may also be universal. Also, we note that the overlap $A_{\bo\bo\bo}$ in the vertex configuration decays as $N^{-5c/36}$, which can be shown in a similar way as Ref.~\cite{zou2021universal}.

\begin{table}[htb!]
    \centering
    \begin{tabular}{p{2cm}|p{2cm}}
    \hline
       overlap  & $p$ \\
         \hline
    $\bo\times \bo\ra\epsilon$     & 0.6656 \\
    $\bo\times \epsilon\ra\bo$ & 0.6656\\
    $\epsilon\times \epsilon\ra\bo$ & 1.3838\\
    $\epsilon\times \epsilon\ra\epsilon$ & 2.0002\\
    $\bo\times \epsilon\ra\epsilon$ & 1.3838\\
    $\bo\times \sigma\ra\sigma$ & 0.6676\\
    $\sigma\times \epsilon\ra\sigma$ & 0.6608\\
    $\sigma\times \sigma\ra \bo$ & 0.6676\\
    $\sigma\times \sigma\ra\epsilon$ & 0.6608 \\
    \hline
    \end{tabular}
    \caption{Finite-size correction of wavefunction overlap in vertex-state configuration, for system size $N\in[100,500)$.}
    \label{tab:Ising_vertex}
\end{table}

\end{document}